\documentclass[prb,superscriptaddress,twocolumn,notitlepage,10pt]{revtex4-2}
\usepackage{amssymb,amsmath,amsthm,amsfonts,mathtools,mathrsfs,chemformula,physics}
\usepackage{graphicx,caption,subcaption,float,comment,bm,enumitem,cancel,xr-hyper}
\usepackage{soul,xcolor}
\usepackage[colorlinks]{hyperref}
\usepackage{cleveref}

\newcommand{\eq}[1]{Eq.~\hyperref[eq:#1]{(\ref*{eq:#1})}}
\renewcommand{\sec}[1]{\hyperref[sec:#1]{Section~\ref*{sec:#1}}}
\newcommand{\app}[1]{\hyperref[app:#1]{Appendix~\ref*{app:#1}}}
\newcommand{\tab}[1]{\hyperref[tab:#1]{Table~\ref*{tab:#1}}}
\newcommand{\fig}[1]{\hyperref[fig:#1]{Figure~\ref*{fig:#1}}}
\newcommand{\figa}[2]{\hyperref[fig:#1]{Figure~\ref*{fig:#1}#2}}
\newcommand{\figx}[2]{\hyperref[fig:#1]{Figure~\ref*{fig:#1}(#2)}}
\newcommand{\thm}[1]{\hyperref[thm:#1]{Theorem~\ref*{thm:#1}}}
\newcommand{\lem}[1]{\hyperref[lem:#1]{Lemma~\ref*{lem:#1}}}
\newcommand{\cor}[1]{\hyperref[cor:#1]{Corollary~\ref*{cor:#1}}}
\newcommand{\defn}[1]{\hyperref[def:#1]{Definition~\ref*{def:#1}}}
\newcommand{\alg}[1]{\hyperref[alg:#1]{Algorithm~\ref*{alg:#1}}}

\newcommand{\be}{\begin{equation}}
\newcommand{\ee}{\end{equation}}
\newcommand{\ba}{\begin{eqnarray}}
\newcommand{\ea}{\end{eqnarray}}

\def\bra#1{\mathinner{\langle{#1}|}}
\def\ket#1{\mathinner{|{#1}\rangle}}

\newcommand{\pvec}[1]{\vec{#1}\mkern2mu\vphantom{#1}}

\makeatletter
\renewcommand*\env@matrix[1][\arraystretch]{%
	\edef\arraystretch{#1}%
	\hskip -\arraycolsep
	\let\@ifnextchar\new@ifnextchar
	\array{*\c@MaxMatrixCols c}}
\makeatother
\makeatletter
\newcommand*{\addFileDependency}[1]{
\typeout{(#1)}
\@addtofilelist{#1}
\IfFileExists{#1}{}{\typeout{No file #1.}}
}\makeatother

\newcommand*{\myexternaldocument}[1]{%
\externaldocument{#1}%
\addFileDependency{#1.tex}%
\addFileDependency{#1.aux}%
}

\myexternaldocument{sm}

\newcommand{\Harvard}{\affiliation{%
Department of Chemistry and Chemical Biology, Harvard University, MA, United States}}

\newcommand{\Columbia}{\affiliation{Department of Chemistry, Columbia University, New York, NY, United States}}

\begin{document}
\setstcolor{red}
\title{Competing Generalized Wigner Crystal States in Moir{\'e} Heterostructures}
\date{\today}

\author{Shu Fay Ung}
\email{su2254@columbia.edu}
\Columbia

\author{Joonho Lee}
\email{joonholee@g.harvard.edu}
\Columbia
\Harvard

\author{David R.~Reichman}
\email{drr2103@columbia.edu}
\Columbia

\begin{abstract}
    We present a comprehensive study of generalized Wigner crystals across various filling factors for system sizes up to 162 holes employing Hartree-Fock theory and explicitly correlated wave function approaches.  While we find broad agreement with the behavior observed in experiments and classical Monte Carlo simulations, we highlight the fact that the Hartree-Fock energy landscape appears to be remarkably complex, exhibiting many competing states, both ordered and disordered, separated by energies of a fraction of $\sim$ 1 meV/hole. We demonstrate which of the located states are metastable by performing a stability analysis at the Hartree-Fock level.  Correlated wave function methods furthermore reveal small correlation energies that are nevertheless large enough to tip the balance of state ordering found within Hartree-Fock theory.
\end{abstract}

\maketitle

\section{Introduction} 
 The interest in a novel class of physical systems--so-called moiré systems--has recently gained tremendous momentum in the condensed-matter physics and materials science communities \cite{andrei_marvels_2021}. These systems are created from two=dimensional layers stacked with a small misalignment such that a characteristic long wavelength moiré pattern forms. The associated moiré period parametrizes the kinetic and interaction energy scales in the stacked system, providing a control knob to alter the electronic Hamiltonian. In particular, the interaction strength between charges can be tuned by changing the twist angle or the layer constituents, rendering such systems a versatile platform for studying strong-correlation physics. Experiments on moiré superlattices have revealed a rich phase diagram of strongly-correlated electronic states, including Mott insulators, generalized Wigner crystals (GWCs), stripe phases, and liquid crystals by varying the interaction strength together with the electron filling factor \cite{wang_correlated_2020,xu_correlated_2020,jin_stripe_nodate,li_imaging_2021-1,regan_mott_2020,huang_correlated_2021}.

Despite the large moiré unit cell, the low-energy physics can be described by an emergent single-particle Hamiltonian (\textit{i.e.} the continuum model) sharing the periodicity of the moiré superlattice. We focus on transition metal dichalcogenide (TMD) \emph{heterobilayers}, where carriers near the Fermi level are localized in one layer only and experience a periodic moiré potential due to the spatial variation in the valence-band maximum of the other layer \cite{zhang_interlayer_2017}. Since the valence-band extrema located at the $K$ and $K^\prime$ momentum points are decoupled and exhibit a large spin-splitting caused by spin-orbit interactions, spin indices are locked with valley degrees of freedom, giving rise to an effective two-fold spin degeneracy in the continuum model \cite{wu_hubbard_2018}.

The accurate description of strongly-correlated states, however, requires the treatment of many-body interactions. A natural approach is to incorporate long-range Coulomb interactions into the continuum model description (\textit{i.e.} the \emph{interacting} continuum model) \cite{zhang_moire_2020,zhang_density_2020,hu_competing_2021}. This framework resembles that of the two-dimensional electron gas (2DEG) with a pinning moiré potential, which may harbor exotic physics arising from Coulomb frustration \cite{jamei_universal_2005} and glassy dynamics \cite{schmalian_stripe_2000} in addition to well-known 2DEG phenomena such as standard Wigner crystallization \cite{tanatar_ground_1989,drummond_phase_2009,loos_uniform_2016}.

Previous theoretical work on GWCs have employed Hartree-Fock (HF) \cite{hu_competing_2021,pan_quantum_2020,pan_interaction-driven_2021,pan_interaction_2022,kaushal_magnetic_2022}, density functional theory \cite{yang_metal-insulator_2023}, classical and quantum Monte Carlo (MC) \cite{padhi_generalized_2021,zhang_electronic_2021,matty_melting_2022,yang_metal-insulator_2023}, and exact diagonalization (ED) \cite{morales-duran_metal-insulator_2021,morales-duran_magnetism_2022} to probe the ground state phase diagram. While classical MC is reasonable deep within the crystalline regime, a comprehensive picture requires a quantum treatment beyond mean-field theory to capture strong correlation effects. Such studies have largely focused on the magnetic behaviour at filling factors $\nu = 1/3, 2/3,$ and 1. Essentially no work has been done to characterize \emph{charge order} across all relevant filling factors using explicitly quantum mechanical approaches in large systems. 

In this paper, we present a comprehensive quantum mechanical study of GWCs in hole-doped TMD heterobilayers at ten distinct filling factors up to $\nu=1$. We first characterize the quantum energy landscape of HF solutions for systems up to 162 holes, establishing that multiple metastable states (as defined by a positive-definite spectrum of the HF Hessian matrix) generically exist, which is suggestive of glassy behaviour in the system. The states are nearly degenerate in energy and may exhibit ordered or disordered configurations. Next, we verify that the complexity in the HF energy landscape persists upon incorporating electron correlation effects via correlated wave function approaches. Correlation energies were found to be small yet sufficiently large to alter the energetic ordering of the states revealed by Hartree-Fock theory. Most notably, our results suggest that quantum fluctuations can stabilize partially crystalline ground states. We demonstrate that such approaches--which have reasonable scalings with system size--may be used to accurately estimate the correlation energies of GWC states. Hartree atomic units are used throughout the text unless stated otherwise.

\section{Model and methods} 
\subsection{Interacting continuum model}
The low-energy physics of TMD heterobilayers is described by a continuum model constructed from the topmost valence band \cite{wu_hubbard_2018}. Carriers near the Fermi level are localized in one layer only, with the effect of the other layer appearing through local variations in the band gap that correspond to moiré length-scale variations in the local atomic registry \cite{zhang_interlayer_2017}. This manifests in the carriers experiencing a moiré potential $\Delta(\vec{r})$ that is periodic with respect to the superlattice, giving rise to the single-particle moiré Hamiltonian \cite{wu_hubbard_2018}
\begin{align}
    \hat{h}_i &= -\frac{\nabla_i^2}{2m^*} + \Delta(\vec{r}_i), \label{moire_h}
\end{align}
where $m^*$ is the carrier effective mass. To the lowest order in a harmonic expansion, we have
\be
    \Delta(\vec{r}) = 2V_m \sum_{j=1}^{3} \cos{\left(\vec{g}_j \cdot \vec{r} + \phi\right)},
\ee
where $V_m$ is the potential strength, $\phi$ characterizes the shape of $\Delta(\vec{r})$ in the moiré unit cell, and $\vec{g}_j$ are the moiré reciprocal lattice vectors in the first shell which satisfy $C_3$ symmetry. The emergence of $\Delta(\vec{r})$ leads to zone-folding of the monolayer Brillouin zones, yielding smaller moiré Brillouin zones (mBZ) that host flat bands. $V_m$ and $\phi$ are intrinsic material properties that can be obtained from fitting to \emph{ab initio} band structures \cite{wu_hubbard_2018,zhang_moire_2020}. 

Taking the carriers to be holes, we set $m^* = -0.45 m_e$ ($m_e$ is the bare electron mass), $V_m = -15$ meV, and $\phi = 45^\circ$ for \ch{WSe2}/\ch{WS2}. At zero twist, the 4\% lattice mismatch between the \ch{WSe2} and \ch{WS2} monolayers produces a moiré superlattice with a lattice constant of $L_m = 8.3$ nm \cite{zhang_moire_2020}. Spin-valley locking \cite{wu_hubbard_2018} implies an effective two-fold degeneracy in the model, thus we identify the filling factor $\nu = 2$ (2 carriers per moiré unit cell) as ``full-filling". We only examine $\nu \leq 1$ since interactions renormalize bands more strongly with higher charge densities, leading to remote band mixing that demands caution with respect to the single-band continuum model for $\nu > 1$ \cite{morales-duran_magnetism_2022}.

The interacting continuum model is obtained by adding Coulomb interactions
\be
    \hat{H} = \sum_{i=1}^N \hat{h}_i + \frac{1}{2} \sum_{ij=1}^N \frac{1}{\epsilon|\vec{r}_i - \vec{r}_j|}. \label{ICM}
\ee
Here, we used a conventional $1/\epsilon$ screened Coulomb potential with $\epsilon$ the effective dielectric constant, although the effects of gate-screening \cite{pan_quantum_2020,morales-duran_magnetism_2022,matty_melting_2022} and a Rytova-Keldysh-type screening \cite{rytova_screened_1967,keldysh_coulomb_1979} can also be considered (see Appendix \ref{app:screened_coulomb} for a discussion). We take  $\epsilon = 7$ to reflect the dielectric environment of hexagonal boron nitride (hBN) encapsulating layers often used in experiments \cite{laturia_dielectric_2018,xu_correlated_2020,li_imaging_2021-1,jin_stripe_nodate}

\subsection{Generalized Hartree-Fock}
Our study employed the generalized Hartree-Fock approximation, where spin-orbitals $\{\chi_i (\vec{r})\}$ comprising the Slater determinant $\ket{\Psi_\text{HF}}$ possess both spin-up and spin-down character, \textit{i.e.} they retain a 2-component spinor structure
\be
    \chi_i(\vec{r}) = 
    \begin{bmatrix}
        \chi_i^{\uparrow}(\vec{r}) \\[0.5em] \chi_i^{\downarrow}(\vec{r}) \label{ghf_orb}
    \end{bmatrix},
\ee
where $\chi_i^\sigma(\vec{r})$ are spatial orbitals in the spin-$\sigma$ sector. By variationally minimizing the HF energy $E_\text{HF} = \bra{\Psi_\text{HF}} \hat{H} \ket{\Psi_\text{HF}}$ with respect to the orbitals $\{\chi_i\}$, we obtain a set of integro-differential equations that can be self-consistently solved for the optimal $\{\chi_i\}$. We represented the $\chi_i^\sigma$ in a plane-wave basis and performed supercell calculations at the $\Gamma$-point of the mBZ. By imposing periodic boundary conditions only over the supercell, we can naturally obtain solutions that are expected to break the superlattice translational symmetry at fractional filling factors \cite{xu_correlated_2020,li_imaging_2021-1,huang_correlated_2021,zhang_electronic_2021,matty_melting_2022,morales-duran_magnetism_2022}. At each filling factor, we found at least one HF solution starting from solutions of the continuum model with matrix elements in the mixed-spin sectors of the initial density matrix $\mathbf{P}$ (\textit{i.e.} $\mathbf{P}^{\uparrow \downarrow}$ and $\mathbf{P}^{\downarrow \uparrow}$) drawn from a standard Gaussian distribution $\mathcal{N}(0, 1)$, normalized by $10^{-3} \times \max|\mathbf{P}|$. This ``noisy" continuum model guess served to unbias solutions away from collinearity. In addition, we selectively located other solutions via physically motivated initial guesses and determined the stability of each state by diagonalizing the HF Hessian matrix. We considered systems with up to 162 holes and, where possible, performed finite-size analyses as detailed below. Further computational details are available in Appendix \ref{app:comp_details}.

\subsection{Estimating the charge gap} \label{estimating_charge_gap}
The charge gap $\Delta_c$ is defined as
\begin{align}
    \Delta_c &= \lim_{N \to \infty} \Delta_c(N), \\
    \Delta_c(N) &= E(N+1) + E(N-1) - 2E(N) \label{delta_scf} \\ 
    &= -\left[E(N) - E(N+1)\right] + \left[E(N-1) - E(N)\right] \nonumber \\
    &= -E_\text{EA} + E_\text{IP} \label{ip_ea},
\end{align}
where $E(N)$ is the total energy of a $N$-particle system, $E_\text{EA} = E(N) - E(N+1)$ is the electron affinity and $E_\text{IP} = E(N-1) - E(N)$ is the ionization potential. Assuming that the orbitals in the $N$ and $N\pm1$ ground states are identical within Hartree-Fock theory, Koopman's theorem \cite{szabo_modern_1996} allows us to identify
\begin{align}
    E_\text{EA} &= -\varepsilon_\text{LUMO}, \quad E_\text{IP} = -\varepsilon_\text{HOMO},
\end{align}
where $\varepsilon_\text{LUMO}$ ($\varepsilon_\text{HOMO}$) is the lowest unoccupied (highest occupied) molecular orbital. This gives
\begin{align}
    \Delta_c(N) \approx \varepsilon_\text{LUMO} - \varepsilon_\text{HOMO}. \label{band_gap}
\end{align}

In the thermodynamic limit, the approximation in \eqref{band_gap} should become an exact equality since the addition or removal of a single particle will have minimal effect on the ground state orbitals. Evaluating \eqref{band_gap} instead of \eqref{delta_scf} furthermore has the advantage of ensuring $\Delta_c \geq 0$ while being computationally cheaper since it only requires one self-consistent loop. However, finite-size errors inevitably arise in practical calculations performed on a finite number $N$ of particles. While periodic boundary conditions help to reduce finite-size effects, non-negligible errors originating from spurious ``self-interactions" between particles and their periodic images still remain--especially for singular long-ranged potentials such as the Coulomb interaction \cite{xing_unified_2021}.

These contributions can be mitigated by employing corrections involving the triangular lattice Madelung energy, given by \cite{bonsall_static_1977}
\be
    \varepsilon_M = \frac{-3.921034}{2 \epsilon \sqrt{\pi}} r_s^{-1} N^{-1/2}, \label{madelung}
\ee
where $\epsilon$ and $r_s$ are the effective dielectric constant and the Wigner-Seitz radius, respectively. For a $N$-particle system, the Madelung-corrected charge gap is \cite{yang_electronic_2020,xing_unified_2021}
\begin{align}
    \tilde{\Delta}_c(N) &= \Delta_c(N) + 2|\varepsilon_M| \label{fs_corrected_charge_gap},
\end{align}
which improves the scaling in the finite-size errors from $\mathcal{O}(N^{-1/2})$ to $\mathcal{O}(N^{-3/2})$ \cite{hunt_quantum_2018}. Given the above considerations, we estimated $\Delta_c$ in the thermodynamic limit by first computing the Madelung-corrected charge gap \eqref{fs_corrected_charge_gap} using \eqref{band_gap}, then extrapolating--where appropriate--$\tilde{\Delta}_c(N)$ to $N \to \infty$ assuming the extrapolation formula
\begin{align}
    \tilde{\Delta}_c(N) = \frac{\gamma}{N^{3/2}} + \Delta_c(\infty),
\end{align}
with $\gamma$ being a parameter to be fitted. Illustrative examples are provided in Appendix \ref{app:finite-size errors}.

\subsection{Correlated methods}
To go beyond mean-field theory and incorporate electron correlation effects, we leveraged several state-of-the-art approaches infrequently used in the study of moiré systems. We investigated systems smaller than our largest HF calculations yet too large to examine by ED--a compromise between system sizes where finite-size errors obscure true correlation effects and computational costs limit the availability of reference solutions for comparison. To find (near) exact energies from our distinct HF solutions, we employed the heat-bath configuration interaction (HCI) \cite{holmes_heat-bath_2016,sharma_semistochastic_2017} and coupled-cluster singles-doubles (CCSD) plus its improvement with perturbative triples (CCSD(T)) approaches \cite{coester_short-range_1960,cizek_correlation_1966,cizek_correlation_1971}. Note that the scaling of CCSD and perhaps CCSD(T) enables their application to the largest systems we examined, making them attractive tools for future investigations of correlation effects in moiré systems \cite{faulstich_interacting_2022}. A brief discussion of these approaches can be found in Appendix \ref{app:corr_methods}.

\section{Results}
\subsection{Hartree-Fock survey across $\nu$} 
We first present a survey of the behavior of HF solutions across $\nu$ for the interacting continuum model. The system sizes we considered are such that the GWCs are commensurate with the supercell and are listed in Table \ref{tab:fs_extrap}. All solutions discussed in this section were initiated from the continuum model guess as described previously and found to be stable.

\begin{table}[h]
    \begin{tabular}{c|l}
        \hline\hline
        Filling factor $\nu$ & System size $N$ \\
        \hline
        1/7 & 7, 28 \\
        1/4 & 4, 16, 36, 49, 64, 81 \\
        1/3 & 3, 12, 27, 48, 75, 108 \\
        2/5 & 10, 40 \\
        1/2 & 8, 18, 32, 50, 72, 98, 162 \\
        3/5 & 15, 60 \\
        2/3 & 6, 24, 54, 96, 150 \\
        3/4 & 12, 27, 48, 75, 108, 147 \\
        6/7 & 42, 168 \\
          1 & 9, 16, 36, 64, 81, 144 \\
        \hline\hline
    \end{tabular}
    \caption{The filling factors and system sizes studied with HF.}
    \label{tab:fs_extrap}
\end{table}

\begin{figure}[t]
    \centering
    \includegraphics[width=0.48\textwidth]{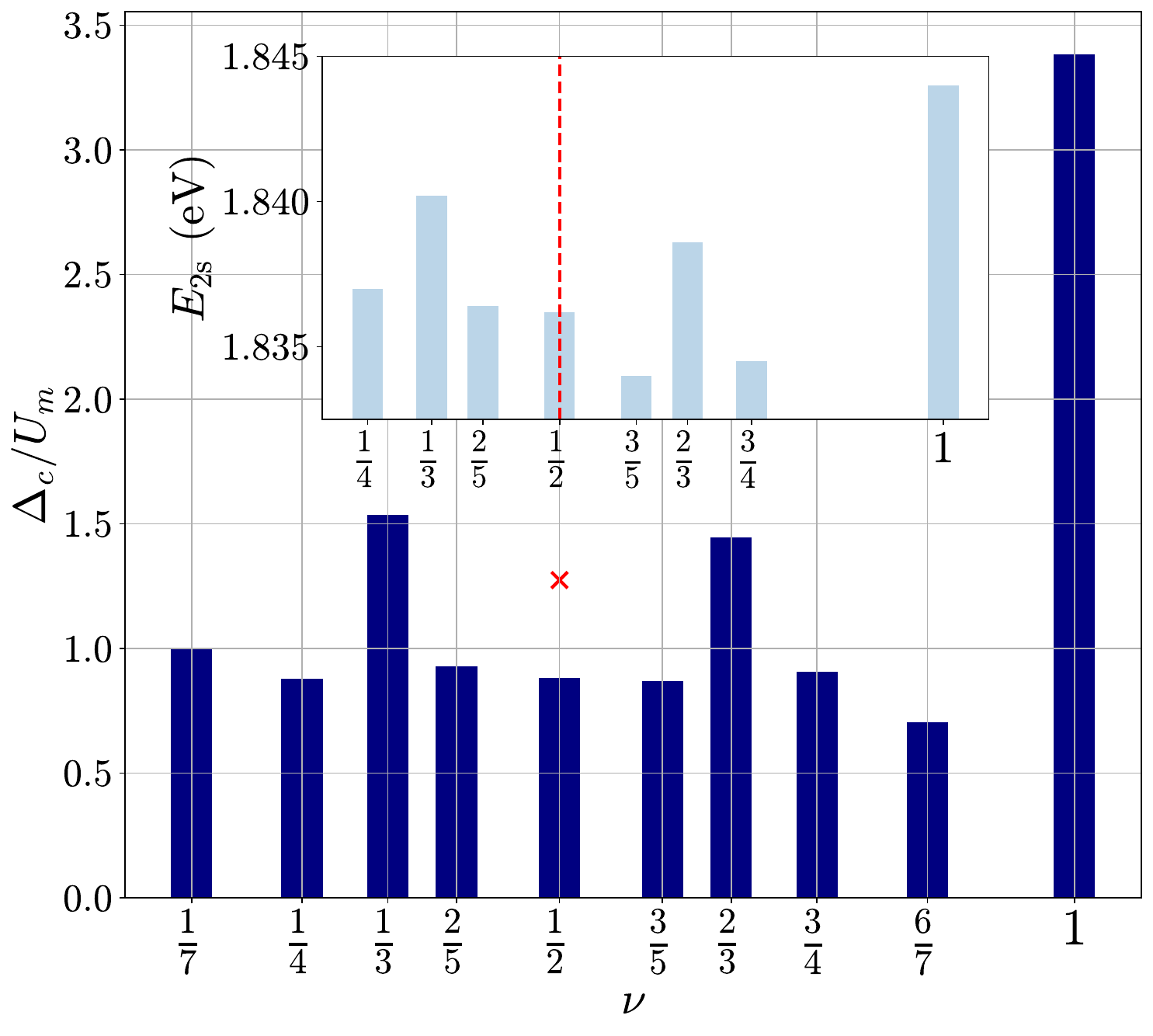} 
    \caption{Charge gaps $\Delta_c$ relative to the moiré scale interaction strength $U_m = 1/\epsilon L_m$ at various fractional filling factors $\nu$, where $L_m$ is the moiré superlattice constant. The inset shows experimental measurements of the 2s exciton resonance energy $E_\text{2s}$ used to probe the charge gap in untwisted \ch{WSe2}/\ch{WS2} heterobilayers \cite{xu_correlated_2020}. All solutions depicted were obtained using the continuum model guess as discussed in the main text, except for the red cross which marks $\Delta_c/U_m$ calculated from the ordered stripe solution.}
    \label{fig:charge_gap}
\end{figure}

Fig. \ref{fig:charge_gap}a shows charge gaps $\Delta_c$ obtained. At $\nu = 1/4, \ 1/3, \ 1/2, \ 2/3, \ 3/4$, and $1$ where there are sufficient data points, we performed the extrapolation as detailed in Section \ref{estimating_charge_gap}. At $\nu = 1/7, \ 2/5, \ 3/5,$ and $6/7$ where there are only two data points, we estimated $\Delta_c(\infty)$ using $\tilde{\Delta}_c(N)$ computed with the largest system size. A large enough $N_\text{basis}^\text{max}$ was used such that all $\tilde{\Delta}_c(N)$ are converged to $<0.5$ meV. At each $\nu$, we also only considered $\Delta_c$ obtained from ordered solutions if they were found; the charge gaps otherwise originate from disordered solutions. Larger values of $\Delta_c$ can occur in ordered solutions (indicated at $\nu = 1/2$ in Fig. \ref{fig:charge_gap}a), but require tuning of the initial guess to find.

\begin{figure*}[phtb]
    \centering
    \includegraphics[width=0.99\textwidth]{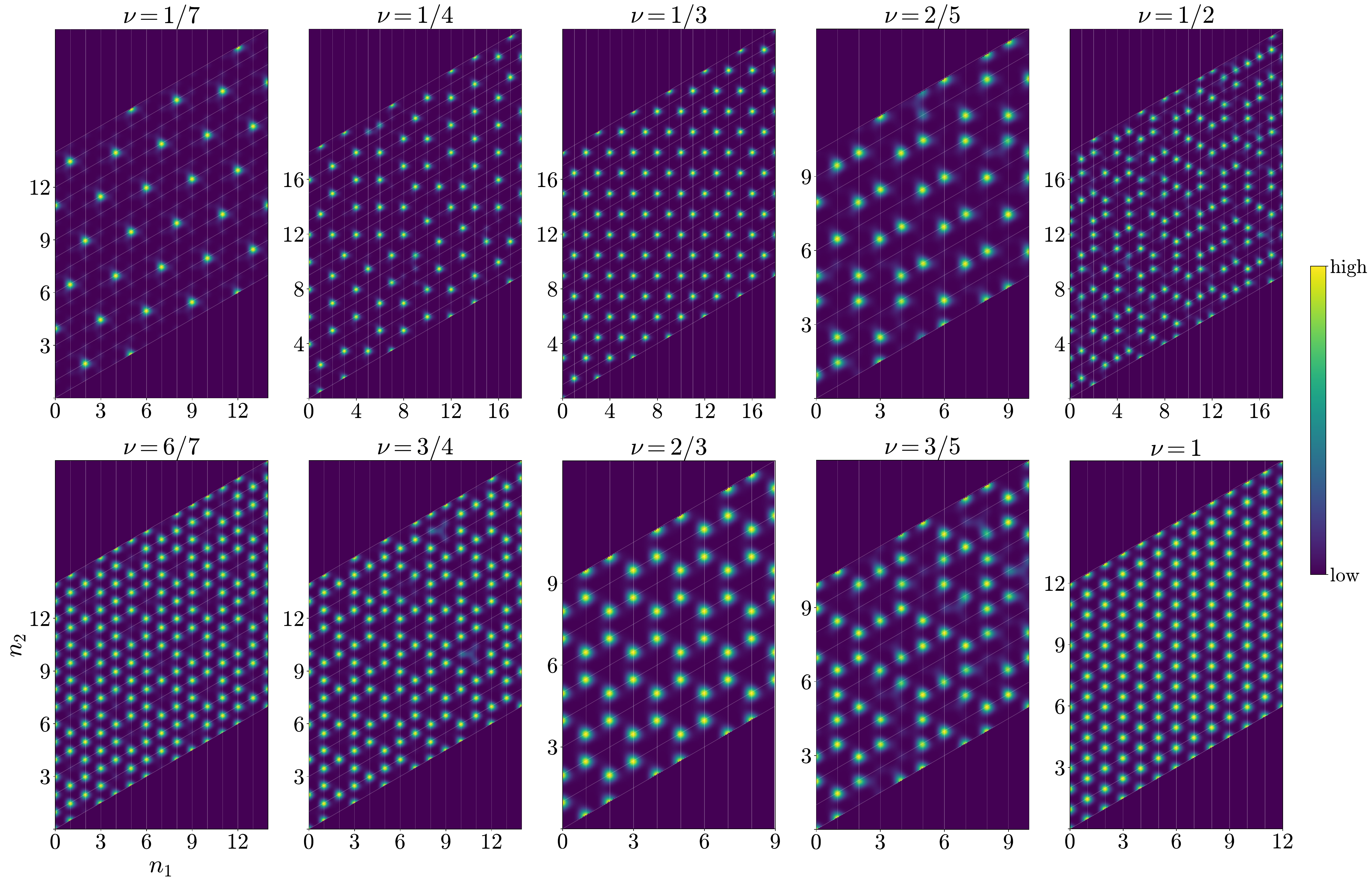}
    \caption{The charge density $\rho(\vec{r})$ at filling factors $\nu \leq 1$ obtained from the unpolarized continuum model guess. The densities are plotted over the supercell and the grid lines show the moiré unit cell boundaries. The indices $n_1, n_2 \in \mathbb{Z}^+$ represent a point $\vec{r} = n_1 \vec{L}_{m,1} + n_2 \vec{L}_{m,2}$ in the supercell, where $\vec{L}_{m,i}$ are the primitive moiré lattice vectors. Disordered configurations were generally observed across $\nu$; ordered phases were seen at $\nu = 1/7,1/3,2/3$, and 1. At $\nu = 1/2$, the continuum model guess yielded a disordered phase in contrast to the ordered phase obtained from the stripe guess (Fig. \ref{fig:charge_density_main}a).}
    \label{fig:charge_density}
\end{figure*}
\begin{figure*}[phbt]
    \centering
    \includegraphics[width=\textwidth]{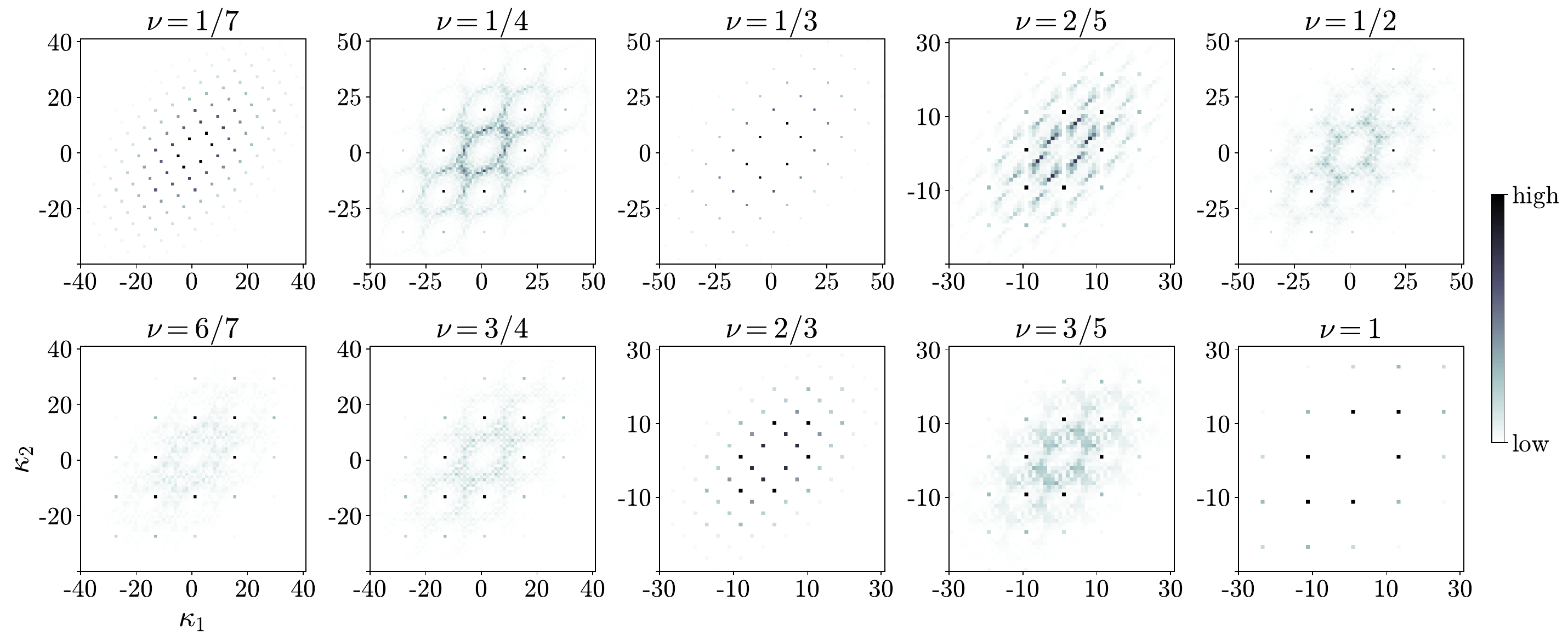}
    \caption{Fast Fourier transform (FFT) of the charge densities in Fig. \ref{fig:charge_density} with the average density subtracted, \textit{i.e.} $|\rho(\vec{k})| = |\mathrm{FFT}\left\{ \rho(\vec{r}) - \langle\rho\rangle \right\}|$. The indices $\kappa_1, \kappa_2 \in \mathbb{Z}$ represent a point $\vec{g} = \kappa_1 \vec{G}_{1} + \kappa_2 \vec{G}_{2}$ in reciprocal space, where $\vec{G}_{i}$ are the primitive reciprocal lattice vectors associated with the supercell.}
    \label{fig:raw_fft_peaks}
\end{figure*}

We plot $\Delta_c$ normalized by the moiré scale interaction strength $U_m = 1/\epsilon L_m$. As in previous experiments \cite{xu_correlated_2020,huang_correlated_2021} and theoretical work \cite{morales-duran_magnetism_2022}, we observed an asymmetry in $\Delta_c$ about $\nu$ = 1/2 at the level of Hartree-Fock theory. The charge gaps for $\nu < 1/2$ are larger than for $\nu > 1/2$ due to the larger Coulomb-to-kinetic-energy ratio at lower average charge densities, which is reminiscent of the low-density behaviour in the 2DEG.  

Compared to the exact diagonalization results of Morales-Durán \textit{et. al.} \cite{morales-duran_magnetism_2022} obtained with small system sizes ($N \leq 16$ for $\nu \leq 1$), we found larger charge gaps and more stable GWCs. Our $(L_m, m^*, V_m)$ parameter values give $W_m/V_m = 0.16$, where $W_m = 1/m^* L_m^2$ is the moiré scale kinetic energy. While they predicted metallic phases across fractional filling factors at this value of $W_m/V_m$, we found insulating states with larger $\Delta_c/U_m$ values than those computed in their insulating states. The discrepancy with the charge gaps reported in \cite{morales-duran_magnetism_2022} appears not to arise entirely due to the use of Hartree-Fock, but due to differences in the model (\textit{i.e.} whether a projection onto continuum model orbitals was involved), the system size, the basis set size, and the manner in which the charge gap was calculated (\textit{e.g.} via \eqref{delta_scf} or \eqref{band_gap}) as well. Replicating the ED calculations of \cite{morales-duran_magnetism_2022}, we found a closing of the gap computed using \eqref{delta_scf} towards metallic behavior with increasing $W_m/V_m$ (\textit{i.e.} decreasing $V_m$) compared to the values reported in Fig. \ref{fig:charge_gap}.

Across all $\nu$, the HF solutions initialized from the continuum model guess consist of localized charge densities, predominantly of a disordered nature. Figs. \ref{fig:charge_density} and \ref{fig:raw_fft_peaks} show the largest basis set charge densities across filling factors and their Fast Fourier transforms (FFTs). The extraction of FFT peak locations from Fig. \ref{fig:raw_fft_peaks} is detailed in Appendix \ref{app:fft_raw}. The $C_3$ symmetry of the moiré superlattice is generally preserved in Fourier space, implying that the charge densities also obey it (with $\nu = 2/5$ perhaps the exception). The superlattice translational symmetry, on the contrary, can be broken. 

The ordered phases we observed occur at $\nu = 1/7, 1/3, 2/3,$ and 1, which exhibit triangular lattice and honeycomb charge orders as seen in experiment \cite{li_imaging_2021-1} and previous theoretical work \cite{liu_excitonic_2021,huang_correlated_2021,morales-duran_magnetism_2022}. At $\nu = 1/3$ and $2/3$, FFT peaks at $\vec{k}$ where $|\vec{k}|/|\vec{G}_m| = 1/\sqrt{3}$ are observed, reflecting a unit cell defined by the vectors
\begin{align}
    \vec{L}_1 &= 2\vec{L}_{m, 1} - \vec{L}_{m, 2}, \quad
    \vec{L}_2 = \vec{L}_{m, 1} + \vec{L}_{m, 2},
\end{align}
where $\vec{L}_{m,i}$ are the primitive moiré lattice vectors. Such a cell is constructed from occupied (vacant) moiré sites for $\nu = 1/3$ ($2/3$). At $\nu = 1/7$, FFT peaks occur at $\vec{k}$ where $|\vec{k}|/|\vec{G}_m| = 1/\sqrt{7}$, indicating a unit cell defined by the vectors
\begin{align}
    \vec{L}_1 &= 3\vec{L}_{m, 1} - \vec{L}_{m, 2}, \quad
    \vec{L}_2 = \vec{L}_{m, 1} + 2\vec{L}_{m, 2}.
\end{align}
In contrast to the case with $\nu = 1/3$ and $2/3$, the $\nu = 1/7$ and $6/7$ charge densities we obtained are not dual \footnote{In the sense that swapping the occupied and vacant sites in the $\nu=1/7$ configuration gives the $\nu=6/7$ configuration} to each other in Fig. \ref{fig:charge_density} and can also be distinguished in Fourier space (Fig. \ref{fig:raw_fft_peaks}). This is likely because our solution at $\nu = 6/7$ is one of several local minima with similar configurations that are close in energy.

Disordered states at other filling factors also resemble MC predictions, \textit{e.g.} the ``columnar dimer crystal" at $\nu = 2/5$ (Fig. \ref{fig:charge_gap}b) \cite{matty_melting_2022}. Notably, our survey revealed a disordered (labyrinthine) stripe configuration at $\nu = 1/2$, in contrast to the ordered stripe phase often discussed in the literature \footnote{A similar labyrinthine solution was also obtained via classical Monte Carlo simulations in the Extended Data of \cite{huang_correlated_2021}}. The fragility of the ordered stripe state has been noted experimentally by Li \textit{et. al.} \cite{li_imaging_2021-1}, who highlight the sensitivity of this phase to lattice strain and local inhomogeneities. In the absence of such experimental considerations, however, we posit that our disordered solution is one of several nearly degenerate local minima (\textit{i.e.} \emph{stable} solutions) that differ slightly in their real-space configurations. As discussed further below, we can also find low-lying ordered states at filling fractions where the continuum model guess yields disordered configurations. This is accomplished through selecting physically motivated initial guesses.

\subsection{Nearly degenerate solutions at $\nu = 1/2$} 
We now focus on $\nu = 1/2$ to probe the existence of nearly degenerate yet distinct HF solutions by supplying different initial guesses. In an attempt to recover the ordered stripe state, we explicitly constructed a polarized ordered stripe guess in addition to the continuum model guess employed earlier. We first present results obtained with $N = 18$ and $N_\mathrm{basis} = 211$ before discussing larger system and basis set size studies.

\begin{figure}[t]
    \centering
    \includegraphics[width=0.47\textwidth]{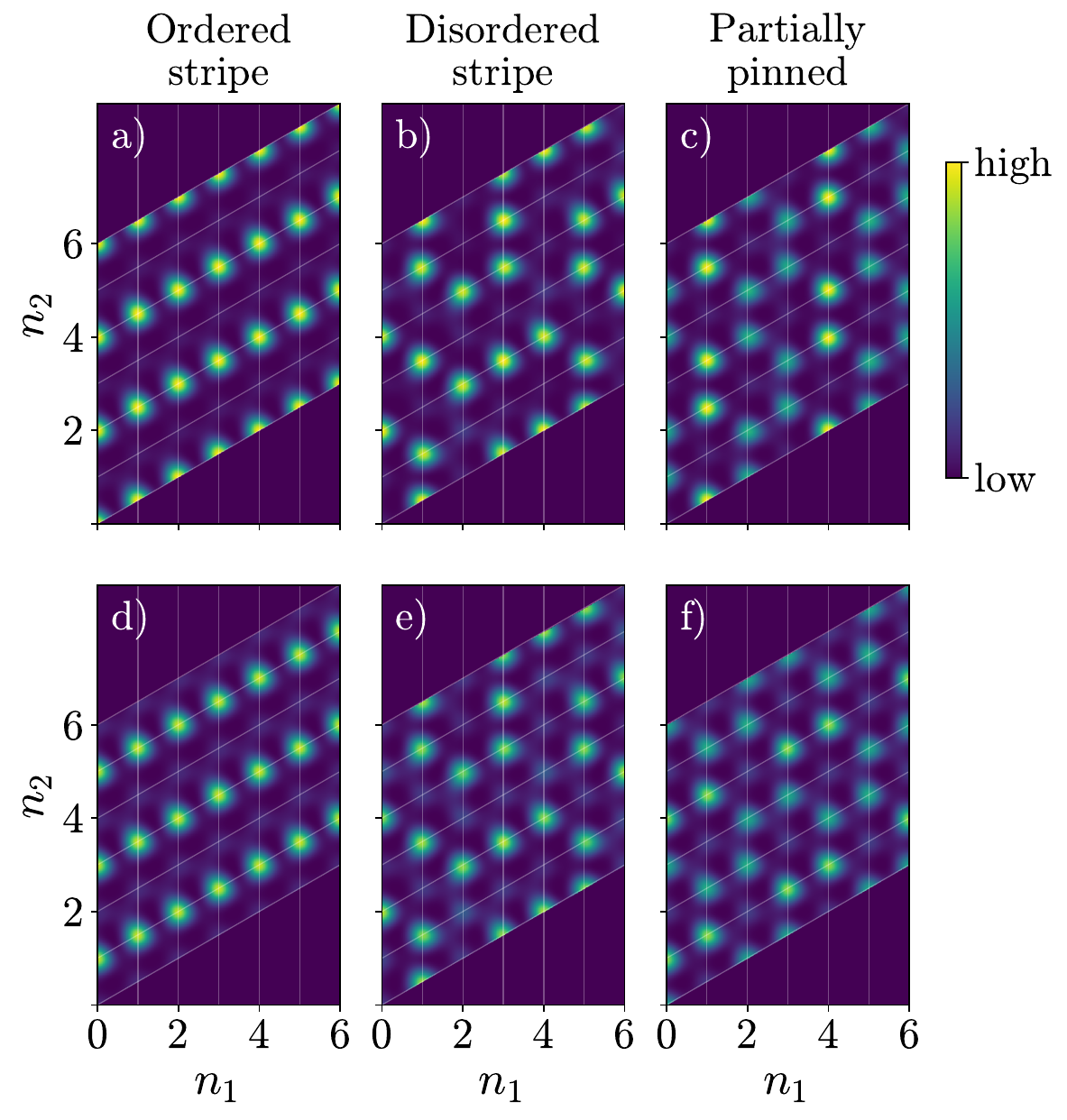} 
    \caption{The charge density $\rho(\vec{r})$ at $\nu = 1/2$. (a)-(c) HF solutions. (d)-(f) CCSD solutions. Comparing the top and bottom rows, we observed that quantum fluctuations induced by correlation delocalizes $\rho(\vec{r})$. Note that $\rho(\vec{r})$ in (a) and (d) are relatively shifted by $L_m$, while (e) and (f) are relatively rotated by $60^\circ$.}
    \label{fig:charge_density_main}
\end{figure}

\begin{figure}[hbt]
    \centering
    \includegraphics[width=0.48\textwidth]{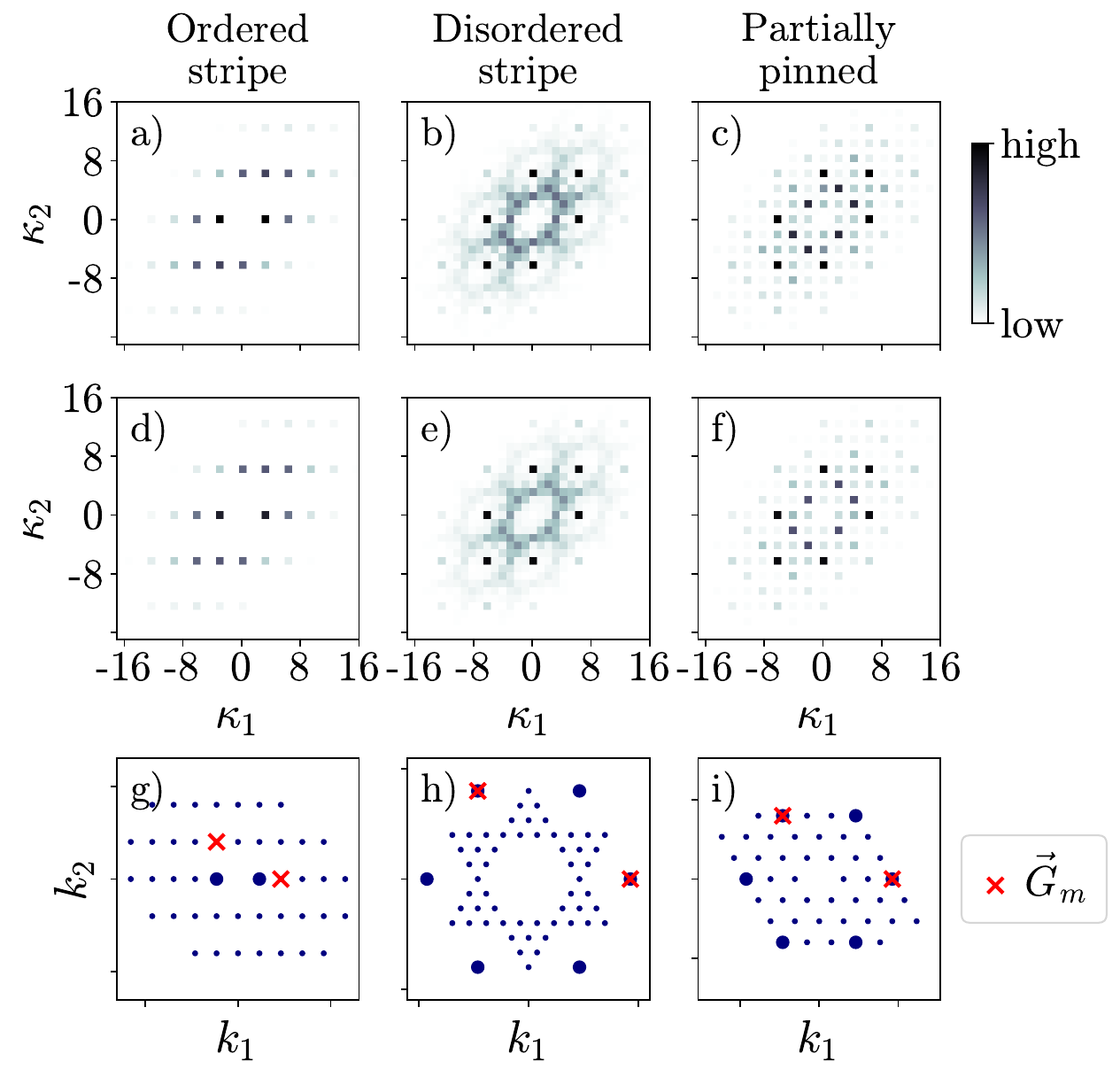} 
    \caption{Fourier-space plots of the charge densities obtained at $\nu = 1/2$. (a)-(f) FFT of the corresponding densities in Fig. \ref{fig:charge_density_main}. (g)-(i) FFT peaks extracted from (a)-(c), plotted directly in reciprocal space $\vec{k} = (k_1, k_2)$. Large dots denote where the FFT signal is maximum and $\vec{G}_m$ are the primitive moiré reciprocal vectors.}
    \label{fig:fft_charge_density_main}
\end{figure}

Figs. \ref{fig:charge_density_main}a-c show the charge configurations obtained from each initialization. Unsurprisingly, the solution from the ordered stripe guess is also an ordered stripe configuration (Fig. \ref{fig:charge_density_main}a), whereas the continuum model guess yielded a disordered (labyrinthine) stripe state (Fig. \ref{fig:charge_density_main}b). Both solutions were verified to be stable in the charge sector. Although several other HF states were discovered throughout our investigation, we only present one additional solution (Fig. \ref{fig:charge_density_main}c) that displays an inhomogeneous density across the honeycomb lattice and is \textit{unstable} (\textit{i.e.} a saddle point) at the HF level. These solutions can be clearly distinguished in Fourier space (Figs. \ref{fig:fft_charge_density_main}a-c)--markedly, the ordered stripe phase exhibits peaks that obey $C_2$ symmetry instead of the $C_3$ symmetry of the moiré superlattice. The finite-size HF energy per particle for these states are $\varepsilon_\mathrm{ordered} = -41.233$ meV, $\varepsilon_\mathrm{disordered} = -41.465$ meV, and $\varepsilon_{\substack{\mathrm{partially}\\\mathrm{pinned}}} = -41.114$ meV, respectively.  Taken together, our results suggest the existence of a rough energy landscape with many closely-lying local minima surrounding the ordered phase, which is reminiscent of a glassy system.

\begin{figure}[htb]
    \centering
    \includegraphics[width=0.48\textwidth]{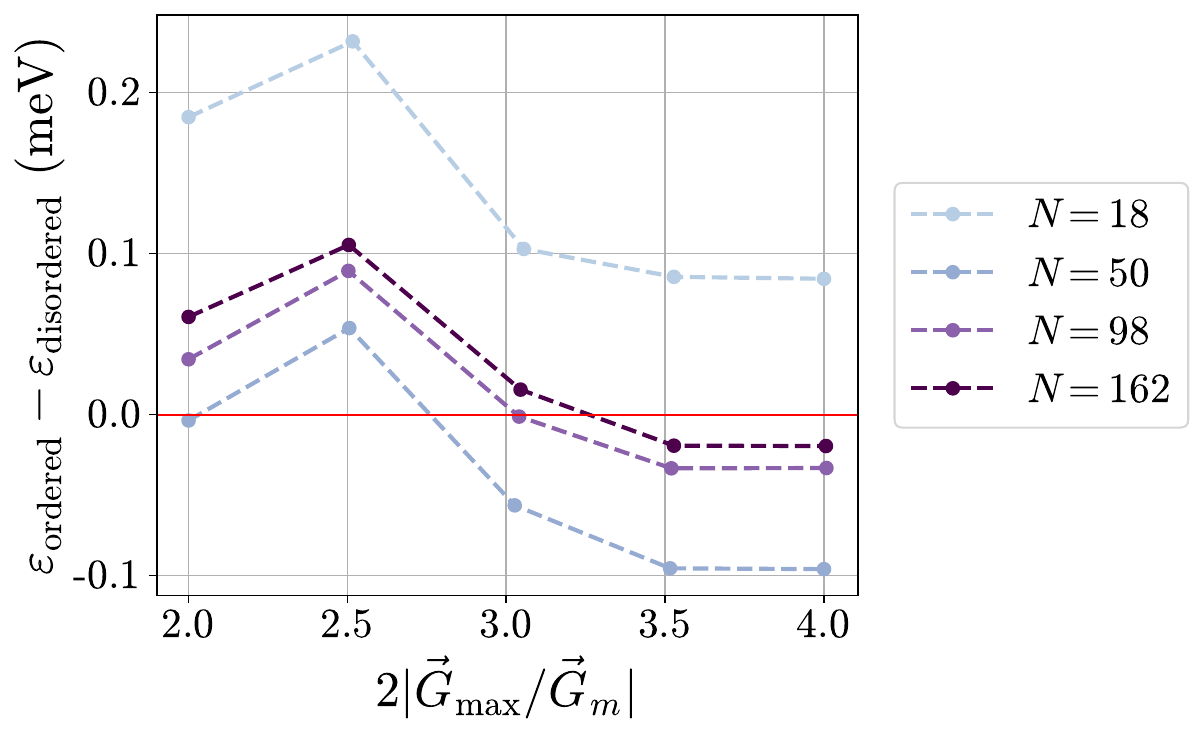}
    \caption{Convergence of the energy difference between the ordered and disordered (labyrinthine) stripe phases with system and basis set size. The quantity  $2|\vec{G}_\mathrm{max} / \vec{G}_m|$ is a measure of the basis set size, where $|\vec{G}_\mathrm{max}|$ is the maximum momentum included in our plane wave basis set. (For reference, $N_\mathrm{basis} = 211$ for $N = 18$ corresponds to $2|\vec{G}_\mathrm{max} / \vec{G}_m| \approx 2.5$.) As $N$ increases, the ordered stripe configuration becomes lower in energy but the difference $|\varepsilon_\mathrm{ordered} - \varepsilon_\mathrm{disordered}|$ decreases.}
    \label{fig:nbsf_conv_nu=1/2_stripe}
\end{figure}

The existence of competing states is furthermore revealed by the convergence of energies with system size. Fig. \ref{fig:nbsf_conv_nu=1/2_stripe} illustrates the convergence of the energy difference per particle $\varepsilon = E(N)/N$ between the ordered and disordered stripe phases with system and basis set size. For the largest $N_\text{basis}$, while the ordered stripe evolves to be the lower energy state as $N$ increases, the two phases also become more nearly degenerate.

\subsection{Further study with correlated methods}
To investigate the effect of quantum fluctuations at $\nu = 1/2$, we performed correlated calculations starting from the mean-field solutions found previously. Again considering the case $N = 18$ and $N_\mathrm{basis} = 211$, we leveraged only spin-collinear variants of HCI, CCSD, and CCSD(T) since the HF solutions we obtained were spin-collinear. We will use ``vHCI" and ``vHCI + PT2" to distinguish between the variational component of HCI and the full algorithm (variational plus perturbative components).

\begin{figure}[t]
    \vspace{2ex}
    \centering
    \includegraphics[width=0.48\textwidth]{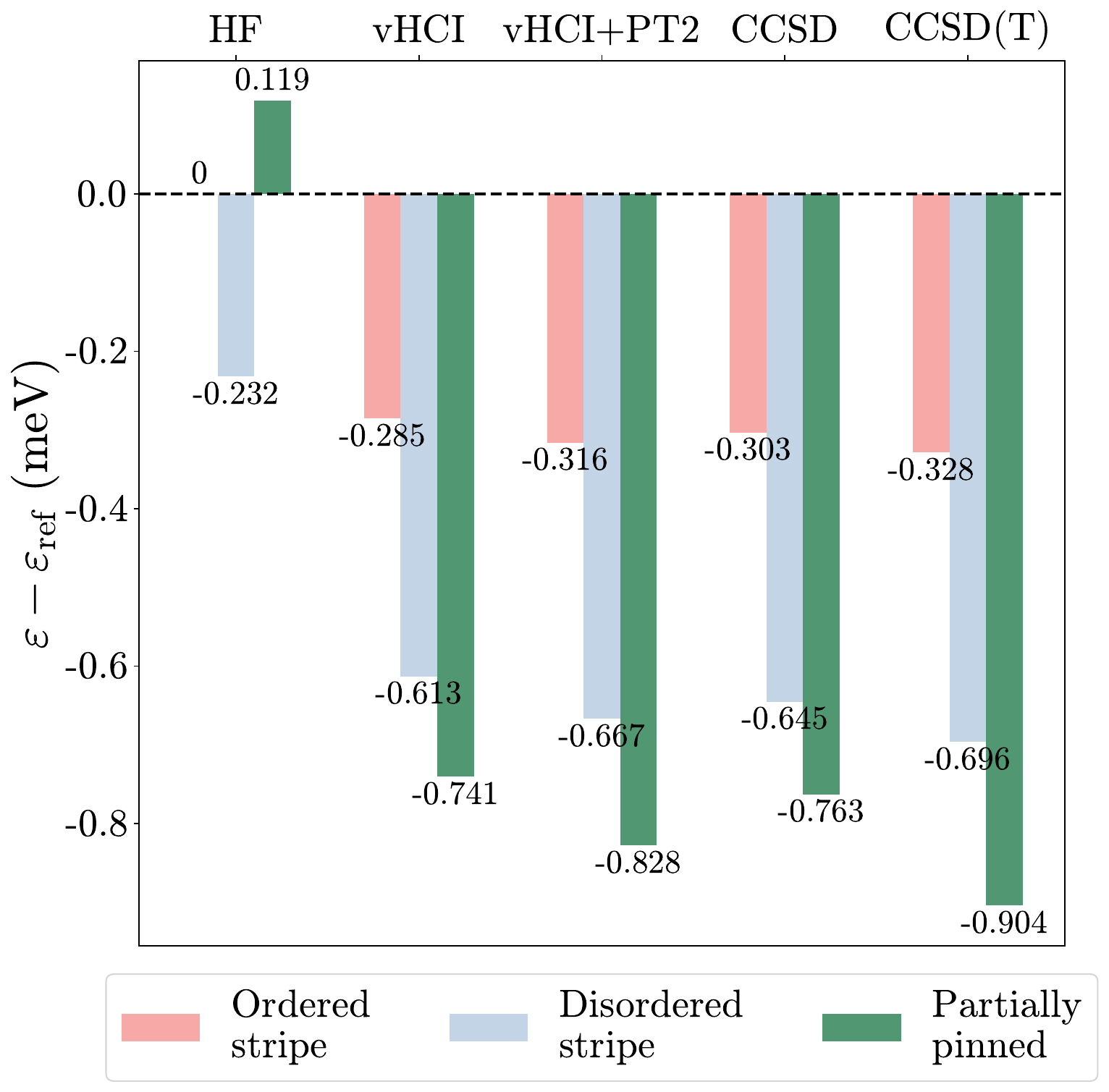} 
    \caption{Energy per particle $\varepsilon$ relative to the HF ordered stripe solution, denoted with $\varepsilon_\mathrm{ref}$. We use vHCI to denote only the variational component of HCI, while vHCI + PT2 denotes the full HCI algorithm. The energy differences are $<1$ meV across all methods. Whereas the energetic ordering found by HF disagrees with correlated approaches, all correlated approaches yield orderings and energies in remarkable agreement with each other.} 
    \label{fig:compare_energy}
\end{figure}

Our results show that the complexity of the energy landscape persists even after introducing electron correlation. Fig. \ref{fig:compare_energy} illustrates the small energy scales associated with the solutions: both the correlation energy and energy differences between the three solutions are $\lesssim 1$ meV. For reference, the relevant energy scales characterizing our system are $U_m \sim V_m \sim \mathcal{O}(10)$ meV and $W_m \sim \mathcal{O}(1)$ meV.  Similar energies obtained across vHCI, vHCI+PT2, CCSD, and CCSD(T) moreover provide confidence that these values are nearly exact. The three correlated solutions remain distinct, which suggests that the metastability associated with the stable HF solutions is retained. Each approach also qualitatively predicts charge densities that agree with each other and HF: panels (d)-(f) of Figs. \ref{fig:charge_density_main} and \ref{fig:fft_charge_density_main} demonstrate how quantum fluctuations delocalize the charge density but still preserve the essential Fourier-space structure from HF. 

Nevertheless, energetic corrections arising from correlation effects are non-trivial. The minuscule correlation energy is sufficient to alter the energetic ordering of states from HF, despite indicating that a single Slater determinant is adequate in providing a quantitative description of each state. For example, while HF predicts the lowest energy state (among our isolated solutions) to be the disordered stripe, all correlated methods predict it to be the partially pinned solution instead--albeit it was unstable at the HF level. 

\section{Conclusions} 
We have numerically studied GWCs in hole-doped TMD heterobilayers at various fractional filling factors using Hartree-Fock and correlated theories. As a concrete example, we focused on the untwisted \ch{WSe2/WS2} heterobilayer embedded in a hBN dielectric environment, which is an experimentally popular system for probing GWCs. Our results show qualitative agreement with experiment in the asymmetry of charge gaps about $\nu = 1/2$ and in the charge orderings observed, particularly at $\nu = 1/3, 2/3,$ and 1. We moreover discovered multiple potentially disordered solutions that suggest the existence of a rugged energy manifold of nearly degenerate states. To probe this further, we focused on the $\nu = 1/2$ case and found that the energy differences between distinct solutions are $<1$ meV at the mean-field level and beyond. Correlation energies are nevertheless non-trivial as demonstrated by a reversal in the energetic ordering of states between HF and correlated methods. Due to quantum fluctuations introduced by such approaches, HF charge densities become more delocalized while still retaining the essential HF features. In addition, we show for relatively small system sizes that fluctuations can stabilize partially pinned states relative to disordered stripe configurations \cite{mahmoudian_glassy_2015,hotta_filling_2007,hotta_strong_2006,watanabe_novel_2006,kaneko_mean-field_2005,mori_non-stripe_2003}. Future work will be devoted to characterizing these states. Our observations reveal the complexity of the moiré energy landscape--defined by several closely-lying metastable states--in both mean-field and correlated theories.

\begin{acknowledgements}
We acknowledge funding support from NSF CHE--2245592 and thank Y. Zhang for help concerning the implementation of the model and A. Millis for useful discussions. Computations in this paper were primarily executed on the Ginsburg High Performance Computing Cluster at Columbia University. Some calculations were run on the FASRC Cannon cluster supported by the FAS Division of Science Research Computing Group at Harvard University. This work also used Bridges-2 Regular Memory (RM) nodes at the Pittsburgh Supercomputing Center and the Expanse cluster at the San Diego Supercomputer Center through allocation CHE220021 from the Advanced Cyberinfrastructure Coordination Ecosystem: Services \& Support (ACCESS) program, which is supported by National Science Foundation grants \#2138259, \#2138286, \#2138307, \#2137603, and \#2138296.
\end{acknowledgements}

\bibliographystyle{apsrev4-1}
\bibliography{ref_short}

\begin{thebibliography}{63}%
\makeatletter
\providecommand \@ifxundefined [1]{%
 \@ifx{#1\undefined}
}%
\providecommand \@ifnum [1]{%
 \ifnum #1\expandafter \@firstoftwo
 \else \expandafter \@secondoftwo
 \fi
}%
\providecommand \@ifx [1]{%
 \ifx #1\expandafter \@firstoftwo
 \else \expandafter \@secondoftwo
 \fi
}%
\providecommand \natexlab [1]{#1}%
\providecommand \enquote  [1]{``#1''}%
\providecommand \bibnamefont  [1]{#1}%
\providecommand \bibfnamefont [1]{#1}%
\providecommand \citenamefont [1]{#1}%
\providecommand \href@noop [0]{\@secondoftwo}%
\providecommand \href [0]{\begingroup \@sanitize@url \@href}%
\providecommand \@href[1]{\@@startlink{#1}\@@href}%
\providecommand \@@href[1]{\endgroup#1\@@endlink}%
\providecommand \@sanitize@url [0]{\catcode `\\12\catcode `\$12\catcode `\&12\catcode `\#12\catcode `\^12\catcode `\_12\catcode `\%12\relax}%
\providecommand \@@startlink[1]{}%
\providecommand \@@endlink[0]{}%
\providecommand \url  [0]{\begingroup\@sanitize@url \@url }%
\providecommand \@url [1]{\endgroup\@href {#1}{\urlprefix }}%
\providecommand \urlprefix  [0]{URL }%
\providecommand \Eprint [0]{\href }%
\providecommand \doibase [0]{http://dx.doi.org/}%
\providecommand \selectlanguage [0]{\@gobble}%
\providecommand \bibinfo  [0]{\@secondoftwo}%
\providecommand \bibfield  [0]{\@secondoftwo}%
\providecommand \translation [1]{[#1]}%
\providecommand \BibitemOpen [0]{}%
\providecommand \bibitemStop [0]{}%
\providecommand \bibitemNoStop [0]{.\EOS\space}%
\providecommand \EOS [0]{\spacefactor3000\relax}%
\providecommand \BibitemShut  [1]{\csname bibitem#1\endcsname}%
\let\auto@bib@innerbib\@empty
\bibitem [{\citenamefont {Andrei}\ \emph {et~al.}(2021)\citenamefont {Andrei}, \citenamefont {Efetov}, \citenamefont {Jarillo-Herrero}, \citenamefont {MacDonald}, \citenamefont {Mak}, \citenamefont {Senthil}, \citenamefont {Tutuc}, \citenamefont {Yazdani},\ and\ \citenamefont {Young}}]{andrei_marvels_2021}%
  \BibitemOpen
  \bibfield  {author} {\bibinfo {author} {\bibfnamefont {E.~Y.}\ \bibnamefont {Andrei}}, \bibinfo {author} {\bibfnamefont {D.~K.}\ \bibnamefont {Efetov}}, \bibinfo {author} {\bibfnamefont {P.}~\bibnamefont {Jarillo-Herrero}}, \bibinfo {author} {\bibfnamefont {A.~H.}\ \bibnamefont {MacDonald}}, \bibinfo {author} {\bibfnamefont {K.~F.}\ \bibnamefont {Mak}}, \bibinfo {author} {\bibfnamefont {T.}~\bibnamefont {Senthil}}, \bibinfo {author} {\bibfnamefont {E.}~\bibnamefont {Tutuc}}, \bibinfo {author} {\bibfnamefont {A.}~\bibnamefont {Yazdani}}, \ and\ \bibinfo {author} {\bibfnamefont {A.~F.}\ \bibnamefont {Young}},\ }\href {\doibase 10.1038/s41578-021-00284-1} {\bibfield  {journal} {\bibinfo  {journal} {Nature Reviews Materials}\ }\textbf {\bibinfo {volume} {6}},\ \bibinfo {pages} {201} (\bibinfo {year} {2021})}\BibitemShut {NoStop}%
\bibitem [{\citenamefont {Wang}\ \emph {et~al.}(2020)\citenamefont {Wang}, \citenamefont {Shih}, \citenamefont {Ghiotto}, \citenamefont {Xian}, \citenamefont {Rhodes}, \citenamefont {Tan}, \citenamefont {Claassen}, \citenamefont {Kennes}, \citenamefont {Bai}, \citenamefont {Kim}, \citenamefont {Watanabe}, \citenamefont {Taniguchi}, \citenamefont {Zhu}, \citenamefont {Hone}, \citenamefont {Rubio}, \citenamefont {Pasupathy},\ and\ \citenamefont {Dean}}]{wang_correlated_2020}%
  \BibitemOpen
  \bibfield  {author} {\bibinfo {author} {\bibfnamefont {L.}~\bibnamefont {Wang}}, \bibinfo {author} {\bibfnamefont {E.-M.}\ \bibnamefont {Shih}}, \bibinfo {author} {\bibfnamefont {A.}~\bibnamefont {Ghiotto}}, \bibinfo {author} {\bibfnamefont {L.}~\bibnamefont {Xian}}, \bibinfo {author} {\bibfnamefont {D.~A.}\ \bibnamefont {Rhodes}}, \bibinfo {author} {\bibfnamefont {C.}~\bibnamefont {Tan}}, \bibinfo {author} {\bibfnamefont {M.}~\bibnamefont {Claassen}}, \bibinfo {author} {\bibfnamefont {D.~M.}\ \bibnamefont {Kennes}}, \bibinfo {author} {\bibfnamefont {Y.}~\bibnamefont {Bai}}, \bibinfo {author} {\bibfnamefont {B.}~\bibnamefont {Kim}}, \bibinfo {author} {\bibfnamefont {K.}~\bibnamefont {Watanabe}}, \bibinfo {author} {\bibfnamefont {T.}~\bibnamefont {Taniguchi}}, \bibinfo {author} {\bibfnamefont {X.}~\bibnamefont {Zhu}}, \bibinfo {author} {\bibfnamefont {J.}~\bibnamefont {Hone}}, \bibinfo {author} {\bibfnamefont {A.}~\bibnamefont {Rubio}}, \bibinfo {author} {\bibfnamefont {A.~N.}\ \bibnamefont {Pasupathy}}, \
  and\ \bibinfo {author} {\bibfnamefont {C.~R.}\ \bibnamefont {Dean}},\ }\href {\doibase 10.1038/s41563-020-0708-6} {\bibfield  {journal} {\bibinfo  {journal} {Nature Materials}\ }\textbf {\bibinfo {volume} {19}},\ \bibinfo {pages} {861} (\bibinfo {year} {2020})}\BibitemShut {NoStop}%
\bibitem [{\citenamefont {Xu}\ \emph {et~al.}(2020)\citenamefont {Xu}, \citenamefont {Liu}, \citenamefont {Rhodes}, \citenamefont {Watanabe}, \citenamefont {Taniguchi}, \citenamefont {Hone}, \citenamefont {Elser}, \citenamefont {Mak},\ and\ \citenamefont {Shan}}]{xu_correlated_2020}%
  \BibitemOpen
  \bibfield  {author} {\bibinfo {author} {\bibfnamefont {Y.}~\bibnamefont {Xu}}, \bibinfo {author} {\bibfnamefont {S.}~\bibnamefont {Liu}}, \bibinfo {author} {\bibfnamefont {D.~A.}\ \bibnamefont {Rhodes}}, \bibinfo {author} {\bibfnamefont {K.}~\bibnamefont {Watanabe}}, \bibinfo {author} {\bibfnamefont {T.}~\bibnamefont {Taniguchi}}, \bibinfo {author} {\bibfnamefont {J.}~\bibnamefont {Hone}}, \bibinfo {author} {\bibfnamefont {V.}~\bibnamefont {Elser}}, \bibinfo {author} {\bibfnamefont {K.~F.}\ \bibnamefont {Mak}}, \ and\ \bibinfo {author} {\bibfnamefont {J.}~\bibnamefont {Shan}},\ }\href {\doibase 10.1038/s41586-020-2868-6} {\bibfield  {journal} {\bibinfo  {journal} {Nature}\ }\textbf {\bibinfo {volume} {587}},\ \bibinfo {pages} {214} (\bibinfo {year} {2020})}\BibitemShut {NoStop}%
\bibitem [{\citenamefont {Jin}\ \emph {et~al.}(2021)\citenamefont {Jin}, \citenamefont {Tao}, \citenamefont {Li}, \citenamefont {Xu}, \citenamefont {Tang}, \citenamefont {Zhu}, \citenamefont {Liu}, \citenamefont {Watanabe}, \citenamefont {Taniguchi}, \citenamefont {Hone}, \citenamefont {Fu}, \citenamefont {Shan},\ and\ \citenamefont {Mak}}]{jin_stripe_nodate}%
  \BibitemOpen
  \bibfield  {author} {\bibinfo {author} {\bibfnamefont {C.}~\bibnamefont {Jin}}, \bibinfo {author} {\bibfnamefont {Z.}~\bibnamefont {Tao}}, \bibinfo {author} {\bibfnamefont {T.}~\bibnamefont {Li}}, \bibinfo {author} {\bibfnamefont {Y.}~\bibnamefont {Xu}}, \bibinfo {author} {\bibfnamefont {Y.}~\bibnamefont {Tang}}, \bibinfo {author} {\bibfnamefont {J.}~\bibnamefont {Zhu}}, \bibinfo {author} {\bibfnamefont {S.}~\bibnamefont {Liu}}, \bibinfo {author} {\bibfnamefont {K.}~\bibnamefont {Watanabe}}, \bibinfo {author} {\bibfnamefont {T.}~\bibnamefont {Taniguchi}}, \bibinfo {author} {\bibfnamefont {J.~C.}\ \bibnamefont {Hone}}, \bibinfo {author} {\bibfnamefont {L.}~\bibnamefont {Fu}}, \bibinfo {author} {\bibfnamefont {J.}~\bibnamefont {Shan}}, \ and\ \bibinfo {author} {\bibfnamefont {K.~F.}\ \bibnamefont {Mak}},\ }\href {\doibase 10.1038/s41563-021-00959-8} {\bibfield  {journal} {\bibinfo  {journal} {Nature Materials}\ }\textbf {\bibinfo {volume} {20}},\ \bibinfo {pages} {940} (\bibinfo {year} {2021})}\BibitemShut
  {NoStop}%
\bibitem [{\citenamefont {Li}\ \emph {et~al.}(2021)\citenamefont {Li}, \citenamefont {Li}, \citenamefont {Regan}, \citenamefont {Wang}, \citenamefont {Zhao}, \citenamefont {Kahn}, \citenamefont {Yumigeta}, \citenamefont {Blei}, \citenamefont {Taniguchi}, \citenamefont {Watanabe}, \citenamefont {Tongay}, \citenamefont {Zettl}, \citenamefont {Crommie},\ and\ \citenamefont {Wang}}]{li_imaging_2021-1}%
  \BibitemOpen
  \bibfield  {author} {\bibinfo {author} {\bibfnamefont {H.}~\bibnamefont {Li}}, \bibinfo {author} {\bibfnamefont {S.}~\bibnamefont {Li}}, \bibinfo {author} {\bibfnamefont {E.~C.}\ \bibnamefont {Regan}}, \bibinfo {author} {\bibfnamefont {D.}~\bibnamefont {Wang}}, \bibinfo {author} {\bibfnamefont {W.}~\bibnamefont {Zhao}}, \bibinfo {author} {\bibfnamefont {S.}~\bibnamefont {Kahn}}, \bibinfo {author} {\bibfnamefont {K.}~\bibnamefont {Yumigeta}}, \bibinfo {author} {\bibfnamefont {M.}~\bibnamefont {Blei}}, \bibinfo {author} {\bibfnamefont {T.}~\bibnamefont {Taniguchi}}, \bibinfo {author} {\bibfnamefont {K.}~\bibnamefont {Watanabe}}, \bibinfo {author} {\bibfnamefont {S.}~\bibnamefont {Tongay}}, \bibinfo {author} {\bibfnamefont {A.}~\bibnamefont {Zettl}}, \bibinfo {author} {\bibfnamefont {M.~F.}\ \bibnamefont {Crommie}}, \ and\ \bibinfo {author} {\bibfnamefont {F.}~\bibnamefont {Wang}},\ }\href {\doibase 10.1038/s41586-021-03874-9} {\bibfield  {journal} {\bibinfo  {journal} {Nature}\ }\textbf {\bibinfo {volume}
  {597}},\ \bibinfo {pages} {650} (\bibinfo {year} {2021})}\BibitemShut {NoStop}%
\bibitem [{\citenamefont {Regan}\ \emph {et~al.}(2020)\citenamefont {Regan}, \citenamefont {Wang}, \citenamefont {Jin}, \citenamefont {Iqbal Bakti~Utama}, \citenamefont {Gao}, \citenamefont {Wei}, \citenamefont {Zhao}, \citenamefont {Zhao}, \citenamefont {Zhang}, \citenamefont {Yumigeta}, \citenamefont {Blei}, \citenamefont {Carlström}, \citenamefont {Watanabe}, \citenamefont {Taniguchi}, \citenamefont {Tongay}, \citenamefont {Crommie}, \citenamefont {Zettl},\ and\ \citenamefont {Wang}}]{regan_mott_2020}%
  \BibitemOpen
  \bibfield  {author} {\bibinfo {author} {\bibfnamefont {E.~C.}\ \bibnamefont {Regan}}, \bibinfo {author} {\bibfnamefont {D.}~\bibnamefont {Wang}}, \bibinfo {author} {\bibfnamefont {C.}~\bibnamefont {Jin}}, \bibinfo {author} {\bibfnamefont {M.}~\bibnamefont {Iqbal Bakti~Utama}}, \bibinfo {author} {\bibfnamefont {B.}~\bibnamefont {Gao}}, \bibinfo {author} {\bibfnamefont {X.}~\bibnamefont {Wei}}, \bibinfo {author} {\bibfnamefont {S.}~\bibnamefont {Zhao}}, \bibinfo {author} {\bibfnamefont {W.}~\bibnamefont {Zhao}}, \bibinfo {author} {\bibfnamefont {Z.}~\bibnamefont {Zhang}}, \bibinfo {author} {\bibfnamefont {K.}~\bibnamefont {Yumigeta}}, \bibinfo {author} {\bibfnamefont {M.}~\bibnamefont {Blei}}, \bibinfo {author} {\bibfnamefont {J.~D.}\ \bibnamefont {Carlström}}, \bibinfo {author} {\bibfnamefont {K.}~\bibnamefont {Watanabe}}, \bibinfo {author} {\bibfnamefont {T.}~\bibnamefont {Taniguchi}}, \bibinfo {author} {\bibfnamefont {S.}~\bibnamefont {Tongay}}, \bibinfo {author} {\bibfnamefont {M.}~\bibnamefont
  {Crommie}}, \bibinfo {author} {\bibfnamefont {A.}~\bibnamefont {Zettl}}, \ and\ \bibinfo {author} {\bibfnamefont {F.}~\bibnamefont {Wang}},\ }\href {\doibase 10.1038/s41586-020-2092-4} {\bibfield  {journal} {\bibinfo  {journal} {Nature}\ }\textbf {\bibinfo {volume} {579}},\ \bibinfo {pages} {359} (\bibinfo {year} {2020})}\BibitemShut {NoStop}%
\bibitem [{\citenamefont {Huang}\ \emph {et~al.}(2021)\citenamefont {Huang}, \citenamefont {Wang}, \citenamefont {Miao}, \citenamefont {Wang}, \citenamefont {Li}, \citenamefont {Lian}, \citenamefont {Taniguchi}, \citenamefont {Watanabe}, \citenamefont {Okamoto}, \citenamefont {Xiao}, \citenamefont {Shi},\ and\ \citenamefont {Cui}}]{huang_correlated_2021}%
  \BibitemOpen
  \bibfield  {author} {\bibinfo {author} {\bibfnamefont {X.}~\bibnamefont {Huang}}, \bibinfo {author} {\bibfnamefont {T.}~\bibnamefont {Wang}}, \bibinfo {author} {\bibfnamefont {S.}~\bibnamefont {Miao}}, \bibinfo {author} {\bibfnamefont {C.}~\bibnamefont {Wang}}, \bibinfo {author} {\bibfnamefont {Z.}~\bibnamefont {Li}}, \bibinfo {author} {\bibfnamefont {Z.}~\bibnamefont {Lian}}, \bibinfo {author} {\bibfnamefont {T.}~\bibnamefont {Taniguchi}}, \bibinfo {author} {\bibfnamefont {K.}~\bibnamefont {Watanabe}}, \bibinfo {author} {\bibfnamefont {S.}~\bibnamefont {Okamoto}}, \bibinfo {author} {\bibfnamefont {D.}~\bibnamefont {Xiao}}, \bibinfo {author} {\bibfnamefont {S.-F.}\ \bibnamefont {Shi}}, \ and\ \bibinfo {author} {\bibfnamefont {Y.-T.}\ \bibnamefont {Cui}},\ }\href {\doibase 10.1038/s41567-021-01171-w} {\bibfield  {journal} {\bibinfo  {journal} {Nature Physics}\ }\textbf {\bibinfo {volume} {17}},\ \bibinfo {pages} {715} (\bibinfo {year} {2021})}\BibitemShut {NoStop}%
\bibitem [{\citenamefont {Zhang}\ \emph {et~al.}(2017)\citenamefont {Zhang}, \citenamefont {Chuu}, \citenamefont {Ren}, \citenamefont {Li}, \citenamefont {Li}, \citenamefont {Jin}, \citenamefont {Chou},\ and\ \citenamefont {Shih}}]{zhang_interlayer_2017}%
  \BibitemOpen
  \bibfield  {author} {\bibinfo {author} {\bibfnamefont {C.}~\bibnamefont {Zhang}}, \bibinfo {author} {\bibfnamefont {C.-P.}\ \bibnamefont {Chuu}}, \bibinfo {author} {\bibfnamefont {X.}~\bibnamefont {Ren}}, \bibinfo {author} {\bibfnamefont {M.-Y.}\ \bibnamefont {Li}}, \bibinfo {author} {\bibfnamefont {L.-J.}\ \bibnamefont {Li}}, \bibinfo {author} {\bibfnamefont {C.}~\bibnamefont {Jin}}, \bibinfo {author} {\bibfnamefont {M.-Y.}\ \bibnamefont {Chou}}, \ and\ \bibinfo {author} {\bibfnamefont {C.-K.}\ \bibnamefont {Shih}},\ }\href {\doibase 10.1126/sciadv.1601459} {\bibfield  {journal} {\bibinfo  {journal} {Science Advances}\ }\textbf {\bibinfo {volume} {3}},\ \bibinfo {pages} {e1601459} (\bibinfo {year} {2017})}\BibitemShut {NoStop}%
\bibitem [{\citenamefont {Wu}\ \emph {et~al.}(2018)\citenamefont {Wu}, \citenamefont {Lovorn}, \citenamefont {Tutuc},\ and\ \citenamefont {MacDonald}}]{wu_hubbard_2018}%
  \BibitemOpen
  \bibfield  {author} {\bibinfo {author} {\bibfnamefont {F.}~\bibnamefont {Wu}}, \bibinfo {author} {\bibfnamefont {T.}~\bibnamefont {Lovorn}}, \bibinfo {author} {\bibfnamefont {E.}~\bibnamefont {Tutuc}}, \ and\ \bibinfo {author} {\bibfnamefont {A.~H.}\ \bibnamefont {MacDonald}},\ }\href {\doibase 10.1103/PhysRevLett.121.026402} {\bibfield  {journal} {\bibinfo  {journal} {Physical Review Letters}\ }\textbf {\bibinfo {volume} {121}},\ \bibinfo {pages} {026402} (\bibinfo {year} {2018})},\ \bibinfo {note} {arXiv: 1804.03151}\BibitemShut {NoStop}%
\bibitem [{\citenamefont {Zhang}\ \emph {et~al.}(2020{\natexlab{a}})\citenamefont {Zhang}, \citenamefont {Yuan},\ and\ \citenamefont {Fu}}]{zhang_moire_2020}%
  \BibitemOpen
  \bibfield  {author} {\bibinfo {author} {\bibfnamefont {Y.}~\bibnamefont {Zhang}}, \bibinfo {author} {\bibfnamefont {N.~F.~Q.}\ \bibnamefont {Yuan}}, \ and\ \bibinfo {author} {\bibfnamefont {L.}~\bibnamefont {Fu}},\ }\href {\doibase 10.1103/PhysRevB.102.201115} {\bibfield  {journal} {\bibinfo  {journal} {Physical Review B}\ }\textbf {\bibinfo {volume} {102}},\ \bibinfo {pages} {201115(R)} (\bibinfo {year} {2020}{\natexlab{a}})},\ \bibinfo {note} {arXiv: 1910.14061}\BibitemShut {NoStop}%
\bibitem [{\citenamefont {Zhang}\ \emph {et~al.}(2020{\natexlab{b}})\citenamefont {Zhang}, \citenamefont {Isobe},\ and\ \citenamefont {Fu}}]{zhang_density_2020}%
  \BibitemOpen
  \bibfield  {author} {\bibinfo {author} {\bibfnamefont {Y.}~\bibnamefont {Zhang}}, \bibinfo {author} {\bibfnamefont {H.}~\bibnamefont {Isobe}}, \ and\ \bibinfo {author} {\bibfnamefont {L.}~\bibnamefont {Fu}},\ }\href {http://arxiv.org/abs/2005.04238} {\  (\bibinfo {year} {2020}{\natexlab{b}})},\ \bibinfo {note} {arXiv: 2005.04238}\BibitemShut {NoStop}%
\bibitem [{\citenamefont {Hu}\ and\ \citenamefont {MacDonald}(2021)}]{hu_competing_2021}%
  \BibitemOpen
  \bibfield  {author} {\bibinfo {author} {\bibfnamefont {N.~C.}\ \bibnamefont {Hu}}\ and\ \bibinfo {author} {\bibfnamefont {A.~H.}\ \bibnamefont {MacDonald}},\ }\href {\doibase 10.1103/PhysRevB.104.214403} {\bibfield  {journal} {\bibinfo  {journal} {Physical Review B}\ }\textbf {\bibinfo {volume} {104}},\ \bibinfo {pages} {214403} (\bibinfo {year} {2021})}\BibitemShut {NoStop}%
\bibitem [{\citenamefont {Jamei}\ \emph {et~al.}(2005)\citenamefont {Jamei}, \citenamefont {Kivelson},\ and\ \citenamefont {Spivak}}]{jamei_universal_2005}%
  \BibitemOpen
  \bibfield  {author} {\bibinfo {author} {\bibfnamefont {R.}~\bibnamefont {Jamei}}, \bibinfo {author} {\bibfnamefont {S.}~\bibnamefont {Kivelson}}, \ and\ \bibinfo {author} {\bibfnamefont {B.}~\bibnamefont {Spivak}},\ }\href {\doibase 10.1103/PhysRevLett.94.056805} {\bibfield  {journal} {\bibinfo  {journal} {Physical Review Letters}\ }\textbf {\bibinfo {volume} {94}},\ \bibinfo {pages} {056805} (\bibinfo {year} {2005})}\BibitemShut {NoStop}%
\bibitem [{\citenamefont {Schmalian}\ and\ \citenamefont {Wolynes}(2000)}]{schmalian_stripe_2000}%
  \BibitemOpen
  \bibfield  {author} {\bibinfo {author} {\bibfnamefont {J.}~\bibnamefont {Schmalian}}\ and\ \bibinfo {author} {\bibfnamefont {P.~G.}\ \bibnamefont {Wolynes}},\ }\href {\doibase 10.1103/PhysRevLett.85.836} {\bibfield  {journal} {\bibinfo  {journal} {Physical Review Letters}\ }\textbf {\bibinfo {volume} {85}},\ \bibinfo {pages} {836} (\bibinfo {year} {2000})}\BibitemShut {NoStop}%
\bibitem [{\citenamefont {Tanatar}\ and\ \citenamefont {Ceperley}(1989)}]{tanatar_ground_1989}%
  \BibitemOpen
  \bibfield  {author} {\bibinfo {author} {\bibfnamefont {B.}~\bibnamefont {Tanatar}}\ and\ \bibinfo {author} {\bibfnamefont {D.~M.}\ \bibnamefont {Ceperley}},\ }\href {\doibase 10.1103/PhysRevB.39.5005} {\bibfield  {journal} {\bibinfo  {journal} {Physical Review B}\ }\textbf {\bibinfo {volume} {39}},\ \bibinfo {pages} {5005} (\bibinfo {year} {1989})}\BibitemShut {NoStop}%
\bibitem [{\citenamefont {Drummond}\ and\ \citenamefont {Needs}(2009)}]{drummond_phase_2009}%
  \BibitemOpen
  \bibfield  {author} {\bibinfo {author} {\bibfnamefont {N.~D.}\ \bibnamefont {Drummond}}\ and\ \bibinfo {author} {\bibfnamefont {R.~J.}\ \bibnamefont {Needs}},\ }\href {\doibase 10.1103/PhysRevLett.102.126402} {\bibfield  {journal} {\bibinfo  {journal} {Physical Review Letters}\ }\textbf {\bibinfo {volume} {102}},\ \bibinfo {pages} {126402} (\bibinfo {year} {2009})}\BibitemShut {NoStop}%
\bibitem [{\citenamefont {Loos}\ and\ \citenamefont {Gill}(2016)}]{loos_uniform_2016}%
  \BibitemOpen
  \bibfield  {author} {\bibinfo {author} {\bibfnamefont {P.-F.}\ \bibnamefont {Loos}}\ and\ \bibinfo {author} {\bibfnamefont {P.~M.~W.}\ \bibnamefont {Gill}},\ }\href {\doibase 10.1002/wcms.1257} {\bibfield  {journal} {\bibinfo  {journal} {WIREs Comput Mol Sci}\ }\textbf {\bibinfo {volume} {6}},\ \bibinfo {pages} {410} (\bibinfo {year} {2016})}\BibitemShut {NoStop}%
\bibitem [{\citenamefont {Pan}\ \emph {et~al.}(2020)\citenamefont {Pan}, \citenamefont {Wu},\ and\ \citenamefont {Das~Sarma}}]{pan_quantum_2020}%
  \BibitemOpen
  \bibfield  {author} {\bibinfo {author} {\bibfnamefont {H.}~\bibnamefont {Pan}}, \bibinfo {author} {\bibfnamefont {F.}~\bibnamefont {Wu}}, \ and\ \bibinfo {author} {\bibfnamefont {S.}~\bibnamefont {Das~Sarma}},\ }\href {\doibase 10.1103/PhysRevB.102.201104} {\bibfield  {journal} {\bibinfo  {journal} {Physical Review B}\ }\textbf {\bibinfo {volume} {102}},\ \bibinfo {pages} {201104(R)} (\bibinfo {year} {2020})}\BibitemShut {NoStop}%
\bibitem [{\citenamefont {Pan}\ and\ \citenamefont {Das~Sarma}(2021)}]{pan_interaction-driven_2021}%
  \BibitemOpen
  \bibfield  {author} {\bibinfo {author} {\bibfnamefont {H.}~\bibnamefont {Pan}}\ and\ \bibinfo {author} {\bibfnamefont {S.}~\bibnamefont {Das~Sarma}},\ }\href {\doibase 10.1103/PhysRevLett.127.096802} {\bibfield  {journal} {\bibinfo  {journal} {Physical Review Letters}\ }\textbf {\bibinfo {volume} {127}},\ \bibinfo {pages} {096802} (\bibinfo {year} {2021})}\BibitemShut {NoStop}%
\bibitem [{\citenamefont {Pan}\ and\ \citenamefont {Das~Sarma}(2022)}]{pan_interaction_2022}%
  \BibitemOpen
  \bibfield  {author} {\bibinfo {author} {\bibfnamefont {H.}~\bibnamefont {Pan}}\ and\ \bibinfo {author} {\bibfnamefont {S.}~\bibnamefont {Das~Sarma}},\ }\href {\doibase 10.1103/PhysRevB.105.041109} {\bibfield  {journal} {\bibinfo  {journal} {Physical Review B}\ }\textbf {\bibinfo {volume} {105}},\ \bibinfo {pages} {L041109} (\bibinfo {year} {2022})}\BibitemShut {NoStop}%
\bibitem [{\citenamefont {Kaushal}\ \emph {et~al.}(2022)\citenamefont {Kaushal}, \citenamefont {Morales-Durán}, \citenamefont {MacDonald},\ and\ \citenamefont {Dagotto}}]{kaushal_magnetic_2022}%
  \BibitemOpen
  \bibfield  {author} {\bibinfo {author} {\bibfnamefont {N.}~\bibnamefont {Kaushal}}, \bibinfo {author} {\bibfnamefont {N.}~\bibnamefont {Morales-Durán}}, \bibinfo {author} {\bibfnamefont {A.~H.}\ \bibnamefont {MacDonald}}, \ and\ \bibinfo {author} {\bibfnamefont {E.}~\bibnamefont {Dagotto}},\ }\href {\doibase 10.1038/s42005-022-01065-0} {\bibfield  {journal} {\bibinfo  {journal} {Communications Physics}\ }\textbf {\bibinfo {volume} {5}},\ \bibinfo {pages} {1} (\bibinfo {year} {2022})}\BibitemShut {NoStop}%
\bibitem [{\citenamefont {Yang}\ \emph {et~al.}(2023)\citenamefont {Yang}, \citenamefont {Morales},\ and\ \citenamefont {Zhang}}]{yang_metal-insulator_2023}%
  \BibitemOpen
  \bibfield  {author} {\bibinfo {author} {\bibfnamefont {Y.}~\bibnamefont {Yang}}, \bibinfo {author} {\bibfnamefont {M.}~\bibnamefont {Morales}}, \ and\ \bibinfo {author} {\bibfnamefont {S.}~\bibnamefont {Zhang}},\ }\href {http://arxiv.org/abs/2306.14954} {} (\bibinfo {year} {2023}),\ \bibinfo {note} {arXiv:2306.14954 [cond-mat]}\BibitemShut {NoStop}%
\bibitem [{\citenamefont {Padhi}\ \emph {et~al.}(2021)\citenamefont {Padhi}, \citenamefont {Chitra},\ and\ \citenamefont {Phillips}}]{padhi_generalized_2021}%
  \BibitemOpen
  \bibfield  {author} {\bibinfo {author} {\bibfnamefont {B.}~\bibnamefont {Padhi}}, \bibinfo {author} {\bibfnamefont {R.}~\bibnamefont {Chitra}}, \ and\ \bibinfo {author} {\bibfnamefont {P.~W.}\ \bibnamefont {Phillips}},\ }\href {\doibase 10.1103/PhysRevB.103.125146} {\bibfield  {journal} {\bibinfo  {journal} {Physical Review B}\ }\textbf {\bibinfo {volume} {103}},\ \bibinfo {pages} {125146} (\bibinfo {year} {2021})}\BibitemShut {NoStop}%
\bibitem [{\citenamefont {Zhang}\ \emph {et~al.}(2021)\citenamefont {Zhang}, \citenamefont {Liu},\ and\ \citenamefont {Fu}}]{zhang_electronic_2021}%
  \BibitemOpen
  \bibfield  {author} {\bibinfo {author} {\bibfnamefont {Y.}~\bibnamefont {Zhang}}, \bibinfo {author} {\bibfnamefont {T.}~\bibnamefont {Liu}}, \ and\ \bibinfo {author} {\bibfnamefont {L.}~\bibnamefont {Fu}},\ }\href {\doibase 10.1103/PhysRevB.103.155142} {\bibfield  {journal} {\bibinfo  {journal} {Physical Review B}\ }\textbf {\bibinfo {volume} {103}},\ \bibinfo {pages} {155142} (\bibinfo {year} {2021})}\BibitemShut {NoStop}%
\bibitem [{\citenamefont {Matty}\ and\ \citenamefont {Kim}(2022)}]{matty_melting_2022}%
  \BibitemOpen
  \bibfield  {author} {\bibinfo {author} {\bibfnamefont {M.}~\bibnamefont {Matty}}\ and\ \bibinfo {author} {\bibfnamefont {E.-A.}\ \bibnamefont {Kim}},\ }\href {\doibase 10.1038/s41467-022-34683-x} {\bibfield  {journal} {\bibinfo  {journal} {Nature Communications}\ }\textbf {\bibinfo {volume} {13}},\ \bibinfo {pages} {7098} (\bibinfo {year} {2022})}\BibitemShut {NoStop}%
\bibitem [{\citenamefont {Morales-Duran}\ \emph {et~al.}(2021)\citenamefont {Morales-Duran}, \citenamefont {MacDonald},\ and\ \citenamefont {Potasz}}]{morales-duran_metal-insulator_2021}%
  \BibitemOpen
  \bibfield  {author} {\bibinfo {author} {\bibfnamefont {N.}~\bibnamefont {Morales-Duran}}, \bibinfo {author} {\bibfnamefont {A.~H.}\ \bibnamefont {MacDonald}}, \ and\ \bibinfo {author} {\bibfnamefont {P.}~\bibnamefont {Potasz}},\ }\href {\doibase 10.1103/PhysRevB.103.L241110} {\bibfield  {journal} {\bibinfo  {journal} {Physical Review B}\ }\textbf {\bibinfo {volume} {103}},\ \bibinfo {pages} {L241110} (\bibinfo {year} {2021})},\ \bibinfo {note} {arXiv:2011.13558 [cond-mat]}\BibitemShut {NoStop}%
\bibitem [{\citenamefont {Morales-Dur\'an}\ \emph {et~al.}(2023)\citenamefont {Morales-Dur\'an}, \citenamefont {Potasz},\ and\ \citenamefont {MacDonald}}]{morales-duran_magnetism_2022}%
  \BibitemOpen
  \bibfield  {author} {\bibinfo {author} {\bibfnamefont {N.}~\bibnamefont {Morales-Dur\'an}}, \bibinfo {author} {\bibfnamefont {P.}~\bibnamefont {Potasz}}, \ and\ \bibinfo {author} {\bibfnamefont {A.~H.}\ \bibnamefont {MacDonald}},\ }\href {\doibase 10.1103/PhysRevB.107.235131} {\bibfield  {journal} {\bibinfo  {journal} {Phys. Rev. B}\ }\textbf {\bibinfo {volume} {107}},\ \bibinfo {pages} {235131} (\bibinfo {year} {2023})}\BibitemShut {NoStop}%
\bibitem [{\citenamefont {Rytova}(1967)}]{rytova_screened_1967}%
  \BibitemOpen
  \bibfield  {author} {\bibinfo {author} {\bibfnamefont {N.~S.}\ \bibnamefont {Rytova}},\ }\href {http://vmu.phys.msu.ru/abstract/1967/3/1967-3-030/} {\bibfield  {journal} {\bibinfo  {journal} {Moscow University Physics Bulletin}\ }\textbf {\bibinfo {volume} {3}},\ \bibinfo {pages} {18} (\bibinfo {year} {1967})}\BibitemShut {NoStop}%
\bibitem [{\citenamefont {Keldysh}(1979)}]{keldysh_coulomb_1979}%
  \BibitemOpen
  \bibfield  {author} {\bibinfo {author} {\bibfnamefont {L.~V.}\ \bibnamefont {Keldysh}},\ }\href {http://jetpletters.ru/ps/1458/article_22207.pdf} {\bibfield  {journal} {\bibinfo  {journal} {JETP Letters}\ }\textbf {\bibinfo {volume} {29}},\ \bibinfo {pages} {716} (\bibinfo {year} {1979})}\BibitemShut {NoStop}%
\bibitem [{\citenamefont {Laturia}\ \emph {et~al.}(2018)\citenamefont {Laturia}, \citenamefont {Van~de Put},\ and\ \citenamefont {Vandenberghe}}]{laturia_dielectric_2018}%
  \BibitemOpen
  \bibfield  {author} {\bibinfo {author} {\bibfnamefont {A.}~\bibnamefont {Laturia}}, \bibinfo {author} {\bibfnamefont {M.~L.}\ \bibnamefont {Van~de Put}}, \ and\ \bibinfo {author} {\bibfnamefont {W.~G.}\ \bibnamefont {Vandenberghe}},\ }\href {\doibase 10.1038/s41699-018-0050-x} {\bibfield  {journal} {\bibinfo  {journal} {npj 2D Materials and Applications}\ }\textbf {\bibinfo {volume} {2}},\ \bibinfo {pages} {6} (\bibinfo {year} {2018})}\BibitemShut {NoStop}%
\bibitem [{\citenamefont {Szabo}\ and\ \citenamefont {Ostlund}(1996)}]{szabo_modern_1996}%
  \BibitemOpen
  \bibfield  {author} {\bibinfo {author} {\bibfnamefont {A.}~\bibnamefont {Szabo}}\ and\ \bibinfo {author} {\bibfnamefont {N.~S.}\ \bibnamefont {Ostlund}},\ }\href@noop {} {\emph {\bibinfo {title} {Modern quantum chemistry}}},\ Dover Books on Chemistry\ (\bibinfo  {publisher} {Dover Publications},\ \bibinfo {address} {Mineola, NY},\ \bibinfo {year} {1996})\BibitemShut {NoStop}%
\bibitem [{\citenamefont {Xing}\ \emph {et~al.}(2023)\citenamefont {Xing}, \citenamefont {Li},\ and\ \citenamefont {Lin}}]{xing_unified_2021}%
  \BibitemOpen
  \bibfield  {author} {\bibinfo {author} {\bibfnamefont {X.}~\bibnamefont {Xing}}, \bibinfo {author} {\bibfnamefont {X.}~\bibnamefont {Li}}, \ and\ \bibinfo {author} {\bibfnamefont {L.}~\bibnamefont {Lin}},\ }\href {\doibase 10.1090/mcom/3877} {\bibfield  {journal} {\bibinfo  {journal} {Mathematics of Computation}\ } (\bibinfo {year} {2023}),\ 10.1090/mcom/3877}\BibitemShut {NoStop}%
\bibitem [{\citenamefont {Bonsall}\ and\ \citenamefont {Maradudin}(1977)}]{bonsall_static_1977}%
  \BibitemOpen
  \bibfield  {author} {\bibinfo {author} {\bibfnamefont {L.}~\bibnamefont {Bonsall}}\ and\ \bibinfo {author} {\bibfnamefont {A.~A.}\ \bibnamefont {Maradudin}},\ }\href {\doibase 10.1103/PhysRevB.15.1959} {\bibfield  {journal} {\bibinfo  {journal} {Physical Review B}\ }\textbf {\bibinfo {volume} {15}},\ \bibinfo {pages} {1959} (\bibinfo {year} {1977})}\BibitemShut {NoStop}%
\bibitem [{\citenamefont {Yang}\ \emph {et~al.}(2020)\citenamefont {Yang}, \citenamefont {Gorelov}, \citenamefont {Pierleoni}, \citenamefont {Ceperley},\ and\ \citenamefont {Holzmann}}]{yang_electronic_2020}%
  \BibitemOpen
  \bibfield  {author} {\bibinfo {author} {\bibfnamefont {Y.}~\bibnamefont {Yang}}, \bibinfo {author} {\bibfnamefont {V.}~\bibnamefont {Gorelov}}, \bibinfo {author} {\bibfnamefont {C.}~\bibnamefont {Pierleoni}}, \bibinfo {author} {\bibfnamefont {D.~M.}\ \bibnamefont {Ceperley}}, \ and\ \bibinfo {author} {\bibfnamefont {M.}~\bibnamefont {Holzmann}},\ }\href {\doibase 10.1103/PhysRevB.101.085115} {\bibfield  {journal} {\bibinfo  {journal} {Physical Review B}\ }\textbf {\bibinfo {volume} {101}},\ \bibinfo {pages} {085115} (\bibinfo {year} {2020})}\BibitemShut {NoStop}%
\bibitem [{\citenamefont {Hunt}\ \emph {et~al.}(2018)\citenamefont {Hunt}, \citenamefont {Szyniszewski}, \citenamefont {Prayogo}, \citenamefont {Maezono},\ and\ \citenamefont {Drummond}}]{hunt_quantum_2018}%
  \BibitemOpen
  \bibfield  {author} {\bibinfo {author} {\bibfnamefont {R.~J.}\ \bibnamefont {Hunt}}, \bibinfo {author} {\bibfnamefont {M.}~\bibnamefont {Szyniszewski}}, \bibinfo {author} {\bibfnamefont {G.~I.}\ \bibnamefont {Prayogo}}, \bibinfo {author} {\bibfnamefont {R.}~\bibnamefont {Maezono}}, \ and\ \bibinfo {author} {\bibfnamefont {N.~D.}\ \bibnamefont {Drummond}},\ }\href {\doibase 10.1103/PhysRevB.98.075122} {\bibfield  {journal} {\bibinfo  {journal} {Physical Review B}\ }\textbf {\bibinfo {volume} {98}},\ \bibinfo {pages} {075122} (\bibinfo {year} {2018})}\BibitemShut {NoStop}%
\bibitem [{\citenamefont {Holmes}\ \emph {et~al.}(2016)\citenamefont {Holmes}, \citenamefont {Tubman},\ and\ \citenamefont {Umrigar}}]{holmes_heat-bath_2016}%
  \BibitemOpen
  \bibfield  {author} {\bibinfo {author} {\bibfnamefont {A.~A.}\ \bibnamefont {Holmes}}, \bibinfo {author} {\bibfnamefont {N.~M.}\ \bibnamefont {Tubman}}, \ and\ \bibinfo {author} {\bibfnamefont {C.~J.}\ \bibnamefont {Umrigar}},\ }\href {\doibase 10.1021/acs.jctc.6b00407} {\bibfield  {journal} {\bibinfo  {journal} {Journal of Chemical Theory and Computation}\ }\textbf {\bibinfo {volume} {12}},\ \bibinfo {pages} {3674} (\bibinfo {year} {2016})}\BibitemShut {NoStop}%
\bibitem [{\citenamefont {Sharma}\ \emph {et~al.}(2017)\citenamefont {Sharma}, \citenamefont {Holmes}, \citenamefont {Jeanmairet}, \citenamefont {Alavi},\ and\ \citenamefont {Umrigar}}]{sharma_semistochastic_2017}%
  \BibitemOpen
  \bibfield  {author} {\bibinfo {author} {\bibfnamefont {S.}~\bibnamefont {Sharma}}, \bibinfo {author} {\bibfnamefont {A.~A.}\ \bibnamefont {Holmes}}, \bibinfo {author} {\bibfnamefont {G.}~\bibnamefont {Jeanmairet}}, \bibinfo {author} {\bibfnamefont {A.}~\bibnamefont {Alavi}}, \ and\ \bibinfo {author} {\bibfnamefont {C.~J.}\ \bibnamefont {Umrigar}},\ }\href {\doibase 10.1021/acs.jctc.6b01028} {\bibfield  {journal} {\bibinfo  {journal} {Journal of Chemical Theory and Computation}\ }\textbf {\bibinfo {volume} {13}},\ \bibinfo {pages} {1595} (\bibinfo {year} {2017})}\BibitemShut {NoStop}%
\bibitem [{\citenamefont {Coester}\ and\ \citenamefont {Kümmel}(1960)}]{coester_short-range_1960}%
  \BibitemOpen
  \bibfield  {author} {\bibinfo {author} {\bibfnamefont {F.}~\bibnamefont {Coester}}\ and\ \bibinfo {author} {\bibfnamefont {H.}~\bibnamefont {Kümmel}},\ }\href {\doibase 10.1016/0029-5582(60)90140-1} {\bibfield  {journal} {\bibinfo  {journal} {Nuclear Physics}\ }\textbf {\bibinfo {volume} {17}},\ \bibinfo {pages} {477} (\bibinfo {year} {1960})}\BibitemShut {NoStop}%
\bibitem [{\citenamefont {Čížek}(1966)}]{cizek_correlation_1966}%
  \BibitemOpen
  \bibfield  {author} {\bibinfo {author} {\bibfnamefont {J.}~\bibnamefont {Čížek}},\ }\href {\doibase 10.1063/1.1727484} {\bibfield  {journal} {\bibinfo  {journal} {The Journal of Chemical Physics}\ }\textbf {\bibinfo {volume} {45}},\ \bibinfo {pages} {4256} (\bibinfo {year} {1966})}\BibitemShut {NoStop}%
\bibitem [{\citenamefont {Čižek}\ and\ \citenamefont {Paldus}(1971)}]{cizek_correlation_1971}%
  \BibitemOpen
  \bibfield  {author} {\bibinfo {author} {\bibfnamefont {J.}~\bibnamefont {Čižek}}\ and\ \bibinfo {author} {\bibfnamefont {J.}~\bibnamefont {Paldus}},\ }\href {\doibase 10.1002/qua.560050402} {\bibfield  {journal} {\bibinfo  {journal} {International Journal of Quantum Chemistry}\ }\textbf {\bibinfo {volume} {5}},\ \bibinfo {pages} {359} (\bibinfo {year} {1971})}\BibitemShut {NoStop}%
\bibitem [{\citenamefont {Faulstich}\ \emph {et~al.}(2023)\citenamefont {Faulstich}, \citenamefont {Stubbs}, \citenamefont {Zhu}, \citenamefont {Soejima}, \citenamefont {Dilip}, \citenamefont {Zhai}, \citenamefont {Kim}, \citenamefont {Zaletel}, \citenamefont {Chan},\ and\ \citenamefont {Lin}}]{faulstich_interacting_2022}%
  \BibitemOpen
  \bibfield  {author} {\bibinfo {author} {\bibfnamefont {F.~M.}\ \bibnamefont {Faulstich}}, \bibinfo {author} {\bibfnamefont {K.~D.}\ \bibnamefont {Stubbs}}, \bibinfo {author} {\bibfnamefont {Q.}~\bibnamefont {Zhu}}, \bibinfo {author} {\bibfnamefont {T.}~\bibnamefont {Soejima}}, \bibinfo {author} {\bibfnamefont {R.}~\bibnamefont {Dilip}}, \bibinfo {author} {\bibfnamefont {H.}~\bibnamefont {Zhai}}, \bibinfo {author} {\bibfnamefont {R.}~\bibnamefont {Kim}}, \bibinfo {author} {\bibfnamefont {M.~P.}\ \bibnamefont {Zaletel}}, \bibinfo {author} {\bibfnamefont {G.~K.-L.}\ \bibnamefont {Chan}}, \ and\ \bibinfo {author} {\bibfnamefont {L.}~\bibnamefont {Lin}},\ }\href {\doibase 10.1103/PhysRevB.107.235123} {\bibfield  {journal} {\bibinfo  {journal} {Phys. Rev. B}\ }\textbf {\bibinfo {volume} {107}},\ \bibinfo {pages} {235123} (\bibinfo {year} {2023})}\BibitemShut {NoStop}%
\bibitem [{\citenamefont {Liu}\ \emph {et~al.}(2021)\citenamefont {Liu}, \citenamefont {Taniguchi}, \citenamefont {Watanabe}, \citenamefont {Gabor}, \citenamefont {Cui},\ and\ \citenamefont {Lui}}]{liu_excitonic_2021}%
  \BibitemOpen
  \bibfield  {author} {\bibinfo {author} {\bibfnamefont {E.}~\bibnamefont {Liu}}, \bibinfo {author} {\bibfnamefont {T.}~\bibnamefont {Taniguchi}}, \bibinfo {author} {\bibfnamefont {K.}~\bibnamefont {Watanabe}}, \bibinfo {author} {\bibfnamefont {N.~M.}\ \bibnamefont {Gabor}}, \bibinfo {author} {\bibfnamefont {Y.-T.}\ \bibnamefont {Cui}}, \ and\ \bibinfo {author} {\bibfnamefont {C.~H.}\ \bibnamefont {Lui}},\ }\href {\doibase 10.1103/PhysRevLett.127.037402} {\bibfield  {journal} {\bibinfo  {journal} {Physical Review Letters}\ }\textbf {\bibinfo {volume} {127}},\ \bibinfo {pages} {037402} (\bibinfo {year} {2021})}\BibitemShut {NoStop}%
\bibitem [{Note1()}]{Note1}%
  \BibitemOpen
  \bibinfo {note} {In the sense that swapping the occupied and vacant sites in the $\nu =1/7$ configuration gives the $\nu =6/7$ configuration}\BibitemShut {NoStop}%
\bibitem [{Note2()}]{Note2}%
  \BibitemOpen
  \bibinfo {note} {A similar labyrinthine solution was also obtained via classical Monte Carlo simulations in the Extended Data of \cite {huang_correlated_2021}}\BibitemShut {NoStop}%
\bibitem [{\citenamefont {Mahmoudian}\ \emph {et~al.}(2015)\citenamefont {Mahmoudian}, \citenamefont {Rademaker}, \citenamefont {Ralko}, \citenamefont {Fratini},\ and\ \citenamefont {Dobrosavljevic}}]{mahmoudian_glassy_2015}%
  \BibitemOpen
  \bibfield  {author} {\bibinfo {author} {\bibfnamefont {S.}~\bibnamefont {Mahmoudian}}, \bibinfo {author} {\bibfnamefont {L.}~\bibnamefont {Rademaker}}, \bibinfo {author} {\bibfnamefont {A.}~\bibnamefont {Ralko}}, \bibinfo {author} {\bibfnamefont {S.}~\bibnamefont {Fratini}}, \ and\ \bibinfo {author} {\bibfnamefont {V.}~\bibnamefont {Dobrosavljevic}},\ }\href {\doibase 10.1103/PhysRevLett.115.025701} {\bibfield  {journal} {\bibinfo  {journal} {Physical Review Letters}\ }\textbf {\bibinfo {volume} {115}},\ \bibinfo {pages} {025701} (\bibinfo {year} {2015})}\BibitemShut {NoStop}%
\bibitem [{\citenamefont {Hotta}\ and\ \citenamefont {Furukawa}(2007)}]{hotta_filling_2007}%
  \BibitemOpen
  \bibfield  {author} {\bibinfo {author} {\bibfnamefont {C.}~\bibnamefont {Hotta}}\ and\ \bibinfo {author} {\bibfnamefont {N.}~\bibnamefont {Furukawa}},\ }\href {\doibase 10.1088/0953-8984/19/14/145242} {\bibfield  {journal} {\bibinfo  {journal} {Journal of Physics: Condensed Matter}\ }\textbf {\bibinfo {volume} {19}},\ \bibinfo {pages} {145242} (\bibinfo {year} {2007})}\BibitemShut {NoStop}%
\bibitem [{\citenamefont {Hotta}\ and\ \citenamefont {Furukawa}(2006)}]{hotta_strong_2006}%
  \BibitemOpen
  \bibfield  {author} {\bibinfo {author} {\bibfnamefont {C.}~\bibnamefont {Hotta}}\ and\ \bibinfo {author} {\bibfnamefont {N.}~\bibnamefont {Furukawa}},\ }\href {\doibase 10.1103/PhysRevB.74.193107} {\bibfield  {journal} {\bibinfo  {journal} {Physical Review B}\ }\textbf {\bibinfo {volume} {74}},\ \bibinfo {pages} {193107} (\bibinfo {year} {2006})}\BibitemShut {NoStop}%
\bibitem [{\citenamefont {Watanabe}\ and\ \citenamefont {Ogata}(2006)}]{watanabe_novel_2006}%
  \BibitemOpen
  \bibfield  {author} {\bibinfo {author} {\bibfnamefont {H.}~\bibnamefont {Watanabe}}\ and\ \bibinfo {author} {\bibfnamefont {M.}~\bibnamefont {Ogata}},\ }\href {\doibase 10.1143/JPSJ.75.063702} {\bibfield  {journal} {\bibinfo  {journal} {Journal of the Physical Society of Japan}\ }\textbf {\bibinfo {volume} {75}},\ \bibinfo {pages} {063702} (\bibinfo {year} {2006})}\BibitemShut {NoStop}%
\bibitem [{\citenamefont {Kaneko}\ and\ \citenamefont {Ogata}(2005)}]{kaneko_mean-field_2005}%
  \BibitemOpen
  \bibfield  {author} {\bibinfo {author} {\bibfnamefont {M.}~\bibnamefont {Kaneko}}\ and\ \bibinfo {author} {\bibfnamefont {M.}~\bibnamefont {Ogata}},\ }\href {\doibase 10.1143/JPSJ.75.014710} {\bibfield  {journal} {\bibinfo  {journal} {Journal of the Physical Society of Japan}\ }\textbf {\bibinfo {volume} {75}},\ \bibinfo {pages} {014710} (\bibinfo {year} {2005})}\BibitemShut {NoStop}%
\bibitem [{\citenamefont {Mori}(2003)}]{mori_non-stripe_2003}%
  \BibitemOpen
  \bibfield  {author} {\bibinfo {author} {\bibfnamefont {T.}~\bibnamefont {Mori}},\ }\href {\doibase 10.1143/JPSJ.72.1469} {\bibfield  {journal} {\bibinfo  {journal} {Journal of the Physical Society of Japan}\ }\textbf {\bibinfo {volume} {72}},\ \bibinfo {pages} {1469} (\bibinfo {year} {2003})}\BibitemShut {NoStop}%
\bibitem [{\citenamefont {Shavitt}\ and\ \citenamefont {Bartlett}(2009)}]{shavitt_many-body_2009}%
  \BibitemOpen
  \bibfield  {author} {\bibinfo {author} {\bibfnamefont {I.}~\bibnamefont {Shavitt}}\ and\ \bibinfo {author} {\bibfnamefont {R.~J.}\ \bibnamefont {Bartlett}},\ }\href@noop {} {\emph {\bibinfo {title} {Many-body methods in chemistry and physics: {MBPT} and coupled-cluster theory}}},\ Cambridge molecular science\ (\bibinfo  {publisher} {Cambridge University Press},\ \bibinfo {address} {Cambridge ; New York},\ \bibinfo {year} {2009})\BibitemShut {NoStop}%
\bibitem [{\citenamefont {Goings}\ \emph {et~al.}(2015)\citenamefont {Goings}, \citenamefont {Ding}, \citenamefont {Frisch},\ and\ \citenamefont {Li}}]{goings_stability_2015}%
  \BibitemOpen
  \bibfield  {author} {\bibinfo {author} {\bibfnamefont {J.~J.}\ \bibnamefont {Goings}}, \bibinfo {author} {\bibfnamefont {F.}~\bibnamefont {Ding}}, \bibinfo {author} {\bibfnamefont {M.~J.}\ \bibnamefont {Frisch}}, \ and\ \bibinfo {author} {\bibfnamefont {X.}~\bibnamefont {Li}},\ }\href {\doibase 10.1063/1.4918561} {\bibfield  {journal} {\bibinfo  {journal} {The Journal of Chemical Physics}\ }\textbf {\bibinfo {volume} {142}},\ \bibinfo {pages} {154109} (\bibinfo {year} {2015})}\BibitemShut {NoStop}%
\bibitem [{\citenamefont {Epifanovsky}\ \emph {et~al.}(2021)\citenamefont {Epifanovsky} \emph {et~al.}}]{epifanovsky_software_2021}%
  \BibitemOpen
  \bibfield  {author} {\bibinfo {author} {\bibfnamefont {E.}~\bibnamefont {Epifanovsky}} \emph {et~al.},\ }\href {\doibase 10.1063/5.0055522} {\bibfield  {journal} {\bibinfo  {journal} {The Journal of Chemical Physics}\ }\textbf {\bibinfo {volume} {155}},\ \bibinfo {pages} {084801} (\bibinfo {year} {2021})}\BibitemShut {NoStop}%
\bibitem [{\citenamefont {Pulay}(1980)}]{pulay_convergence_1980}%
  \BibitemOpen
  \bibfield  {author} {\bibinfo {author} {\bibfnamefont {P.}~\bibnamefont {Pulay}},\ }\href {\doibase 10.1016/0009-2614(80)80396-4} {\bibfield  {journal} {\bibinfo  {journal} {Chemical Physics Letters}\ }\textbf {\bibinfo {volume} {73}},\ \bibinfo {pages} {393} (\bibinfo {year} {1980})}\BibitemShut {NoStop}%
\bibitem [{\citenamefont {Pulay}(1982)}]{pulay_improved_1982}%
  \BibitemOpen
  \bibfield  {author} {\bibinfo {author} {\bibfnamefont {P.}~\bibnamefont {Pulay}},\ }\href {\doibase 10.1002/jcc.540030413} {\bibfield  {journal} {\bibinfo  {journal} {Journal of Computational Chemistry}\ }\textbf {\bibinfo {volume} {3}},\ \bibinfo {pages} {556} (\bibinfo {year} {1982})}\BibitemShut {NoStop}%
\bibitem [{\citenamefont {Van~Voorhis}\ and\ \citenamefont {Head-Gordon}(2002)}]{van_voorhis_geometric_2002}%
  \BibitemOpen
  \bibfield  {author} {\bibinfo {author} {\bibfnamefont {T.}~\bibnamefont {Van~Voorhis}}\ and\ \bibinfo {author} {\bibfnamefont {M.}~\bibnamefont {Head-Gordon}},\ }\href {\doibase 10.1080/00268970110103642} {\bibfield  {journal} {\bibinfo  {journal} {Molecular Physics}\ }\textbf {\bibinfo {volume} {100}},\ \bibinfo {pages} {1713} (\bibinfo {year} {2002})}\BibitemShut {NoStop}%
\bibitem [{\citenamefont {Sharada}\ \emph {et~al.}(2015)\citenamefont {Sharada}, \citenamefont {Stück}, \citenamefont {Sundstrom}, \citenamefont {Bell},\ and\ \citenamefont {Head-Gordon}}]{sharada_wavefunction_2015}%
  \BibitemOpen
  \bibfield  {author} {\bibinfo {author} {\bibfnamefont {S.~M.}\ \bibnamefont {Sharada}}, \bibinfo {author} {\bibfnamefont {D.}~\bibnamefont {Stück}}, \bibinfo {author} {\bibfnamefont {E.~J.}\ \bibnamefont {Sundstrom}}, \bibinfo {author} {\bibfnamefont {A.~T.}\ \bibnamefont {Bell}}, \ and\ \bibinfo {author} {\bibfnamefont {M.}~\bibnamefont {Head-Gordon}},\ }\href {\doibase 10.1080/00268976.2015.1014442} {\bibfield  {journal} {\bibinfo  {journal} {Molecular Physics}\ }\textbf {\bibinfo {volume} {113}},\ \bibinfo {pages} {1802} (\bibinfo {year} {2015})}\BibitemShut {NoStop}%
\bibitem [{\citenamefont {Small}\ \emph {et~al.}(2015)\citenamefont {Small}, \citenamefont {Sundstrom},\ and\ \citenamefont {Head-Gordon}}]{small_simple_2015}%
  \BibitemOpen
  \bibfield  {author} {\bibinfo {author} {\bibfnamefont {D.~W.}\ \bibnamefont {Small}}, \bibinfo {author} {\bibfnamefont {E.~J.}\ \bibnamefont {Sundstrom}}, \ and\ \bibinfo {author} {\bibfnamefont {M.}~\bibnamefont {Head-Gordon}},\ }\href {\doibase 10.1063/1.4913740} {\bibfield  {journal} {\bibinfo  {journal} {The Journal of Chemical Physics}\ }\textbf {\bibinfo {volume} {142}},\ \bibinfo {pages} {094112} (\bibinfo {year} {2015})}\BibitemShut {NoStop}%
\bibitem [{\citenamefont {Sun}(2015)}]{sun_libcint_2015}%
  \BibitemOpen
  \bibfield  {author} {\bibinfo {author} {\bibfnamefont {Q.}~\bibnamefont {Sun}},\ }\href {\doibase 10.1002/jcc.23981} {\bibfield  {journal} {\bibinfo  {journal} {Journal of Computational Chemistry}\ }\textbf {\bibinfo {volume} {36}},\ \bibinfo {pages} {1664} (\bibinfo {year} {2015})}\BibitemShut {NoStop}%
\bibitem [{\citenamefont {Sun}\ \emph {et~al.}(2018)\citenamefont {Sun}, \citenamefont {Berkelbach}, \citenamefont {Blunt}, \citenamefont {Booth}, \citenamefont {Guo}, \citenamefont {Li}, \citenamefont {Liu}, \citenamefont {McClain}, \citenamefont {Sayfutyarova}, \citenamefont {Sharma}, \citenamefont {Wouters},\ and\ \citenamefont {Chan}}]{sun_pyscf_2018}%
  \BibitemOpen
  \bibfield  {author} {\bibinfo {author} {\bibfnamefont {Q.}~\bibnamefont {Sun}}, \bibinfo {author} {\bibfnamefont {T.~C.}\ \bibnamefont {Berkelbach}}, \bibinfo {author} {\bibfnamefont {N.~S.}\ \bibnamefont {Blunt}}, \bibinfo {author} {\bibfnamefont {G.~H.}\ \bibnamefont {Booth}}, \bibinfo {author} {\bibfnamefont {S.}~\bibnamefont {Guo}}, \bibinfo {author} {\bibfnamefont {Z.}~\bibnamefont {Li}}, \bibinfo {author} {\bibfnamefont {J.}~\bibnamefont {Liu}}, \bibinfo {author} {\bibfnamefont {J.~D.}\ \bibnamefont {McClain}}, \bibinfo {author} {\bibfnamefont {E.~R.}\ \bibnamefont {Sayfutyarova}}, \bibinfo {author} {\bibfnamefont {S.}~\bibnamefont {Sharma}}, \bibinfo {author} {\bibfnamefont {S.}~\bibnamefont {Wouters}}, \ and\ \bibinfo {author} {\bibfnamefont {G.~K.-L.}\ \bibnamefont {Chan}},\ }\href {\doibase 10.1002/wcms.1340} {\bibfield  {journal} {\bibinfo  {journal} {WIREs Computational Molecular Science}\ }\textbf {\bibinfo {volume} {8}},\ \bibinfo {pages} {e1340} (\bibinfo {year} {2018})}\BibitemShut {NoStop}%
\bibitem [{\citenamefont {Sun}\ \emph {et~al.}(2020)\citenamefont {Sun} \emph {et~al.}}]{sun_recent_2020}%
  \BibitemOpen
  \bibfield  {author} {\bibinfo {author} {\bibfnamefont {Q.}~\bibnamefont {Sun}} \emph {et~al.},\ }\href {\doibase 10.1063/5.0006074} {\bibfield  {journal} {\bibinfo  {journal} {The Journal of Chemical Physics}\ }\textbf {\bibinfo {volume} {153}},\ \bibinfo {pages} {024109} (\bibinfo {year} {2020})}\BibitemShut {NoStop}%
\bibitem [{\citenamefont {Smith}\ \emph {et~al.}(2017)\citenamefont {Smith}, \citenamefont {Mussard}, \citenamefont {Holmes},\ and\ \citenamefont {Sharma}}]{smith_cheap_2017}%
  \BibitemOpen
  \bibfield  {author} {\bibinfo {author} {\bibfnamefont {J.~E.~T.}\ \bibnamefont {Smith}}, \bibinfo {author} {\bibfnamefont {B.}~\bibnamefont {Mussard}}, \bibinfo {author} {\bibfnamefont {A.~A.}\ \bibnamefont {Holmes}}, \ and\ \bibinfo {author} {\bibfnamefont {S.}~\bibnamefont {Sharma}},\ }\href {\doibase 10.1021/acs.jctc.7b00900} {\bibfield  {journal} {\bibinfo  {journal} {Journal of Chemical Theory and Computation}\ }\textbf {\bibinfo {volume} {13}},\ \bibinfo {pages} {5468} (\bibinfo {year} {2017})}\BibitemShut {NoStop}%
\bibitem [{\citenamefont {Bernevig}\ \emph {et~al.}(2021)\citenamefont {Bernevig}, \citenamefont {Song}, \citenamefont {Regnault},\ and\ \citenamefont {Lian}}]{bernevig_twisted_2021}%
  \BibitemOpen
  \bibfield  {author} {\bibinfo {author} {\bibfnamefont {B.~A.}\ \bibnamefont {Bernevig}}, \bibinfo {author} {\bibfnamefont {Z.-D.}\ \bibnamefont {Song}}, \bibinfo {author} {\bibfnamefont {N.}~\bibnamefont {Regnault}}, \ and\ \bibinfo {author} {\bibfnamefont {B.}~\bibnamefont {Lian}},\ }\href {\doibase 10.1103/PhysRevB.103.205413} {\bibfield  {journal} {\bibinfo  {journal} {Physical Review B}\ }\textbf {\bibinfo {volume} {103}},\ \bibinfo {pages} {205413} (\bibinfo {year} {2021})}\BibitemShut {NoStop}%
\end{thebibliography}%

\appendix
\section{Model details} \label{app:model}
The interacting continuum model is described by the Hamiltonian
\begin{align}
    \hat{H} = \sum_{i=1}^N \left\{-\frac{\nabla_i^2}{2m^*} + \Delta(\vec{r}_i) \right\} + \frac{1}{2} \sum_{ij=1}^N \frac{1}{\epsilon|\vec{r}_i - \vec{r}_j|}, \label{app:ICM}
\end{align}
where $m^*$ is the carrier effective mass, $\Delta$ is the moiré potential, and $\epsilon$ is the effective dielectric constant. In our calculations, we instead work with a scaled Hamiltonian $\hat{H}'$ defined by
\begin{align}
    \hat{H}' &= \sum_{i=1}^N \left\{-\frac{\nabla_i^{\prime 2}}{2} + \Delta'(\pvec{r}'_i) \right\} + \frac{1}{2} \sum_{ij=1}^N \frac{1}{|\pvec{r}'_i - \pvec{r}'_j|}, \label{app:scaled_ICM} \\
    \hat{H}' &= \alpha \hat{H}, \qquad \Delta' = \alpha \Delta, \qquad \pvec{r}' = \beta \vec{r}, \\
    \alpha &= \frac{\epsilon^2}{m^*}, \qquad \beta = \frac{m^*}{\epsilon},
\end{align}
which effectively maps the interacting continuum model to a 2DEG in the scaled moiré potential $\Delta'$ \cite{zhang_density_2020}. This form allows the interacting continuum model to be treated with DFT employing exchange-correlation functionals derived from the 2DEG. Note that if $\ket{\Psi'}$ is an eigenstate of $\hat{H}'$ with energy $E'$, then
\begin{align}
    \hat{H}' \ket{\Psi'} = \alpha \hat{H} \ket{\Psi'} &= E' \ket{\Psi'}, \nonumber \\
    \implies \hat{H} \ket{\Psi'} &= \frac{E'}{\alpha} \ket{\Psi'},
\end{align}
\textit{i.e.} $\ket{\Psi'}$ is also an eigenstate of the original Hamiltonian $\hat{H}$ with energy $E = E'/\alpha$. Thus, we divided energies obtained from the scaled Hamiltonian by $\alpha$ to recover energies associated with the physical moiré system.

\section{Correlated methods} \label{app:corr_methods}
Any antisymmetric $N$-particle state $\ket{\Psi}$ can be represented in the basis consisting of a reference Slater determinant $\ket{\Phi}$ and all of its excitations, \textit{i.e.}
\begin{align}
    \ket{\Psi} &= \left( c_0  + \sum_{ia} c_i^a \hat{a}_a^\dagger \hat{a}_i + \sum_{\substack{i<j\\a<b}} c_{ij}^{ab} \hat{a}_a^\dagger \hat{a}_b^\dagger \hat{a}_i \hat{a}_j + \cdots \right) \ket{\Phi}, \label{ci_wfn}
\end{align}
where $c_i^a, c_{ij}^{ab}, \dots$ are coefficients and $\hat{a}_i^\dagger (\hat{a}_i)$ are creation (annihilation) operators defined over the orbital basis set  $\{\phi_i\}$ from which $\ket{\Phi}$ is constructed \cite{szabo_modern_1996}. This representation becomes exact as the orbital basis approaches completeness. In principle, the coefficients can be determined through exact diagonalization, but the problem quickly becomes intractable as the Hilbert space scales exponentially with system size. Correlated wave function methods thus seek to approximate $\eqref{ci_wfn}$ starting from some known reference $\ket{\Phi}$--often obtained from Hartree-Fock theory (hence by \emph{correlation}, we refer to contributions beyond Hartree-Fock). HCI \cite{holmes_heat-bath_2016,sharma_semistochastic_2017} and coupled-cluster (CC) theories \cite{coester_short-range_1960,cizek_correlation_1966,cizek_correlation_1971} are two such methods that we employ in our paper.

\subsection{Heat-bath configuration interaction}
Given a Hamiltonian $\hat{H}$, HCI finds an approximation to the ground state $\ket{\Psi}$ consisting of a selected set of the most important determinants only. This is achieved in two stages: first, a variational wave function is constructed from a set of chosen determinants; second, perturbative corrections to the variational energy determine additional determinants to be added. The selection of determinants in each stage is tuned by two parameters, $\epsilon_1$ and $\epsilon_2$, which control the tradeoff between speed and accuracy. We use vHCI and vHCI + PT2 to denote the variational stage and its correction with perturbation theory (\textit{i.e.} the full HCI algorithm), respectively. 

vHCI consists of iteratively expanding the selected space (which usually only includes the Hartree-Fock determinant initially) while minimizing the selected-space energy. At each iteration, new determinants $\ket{\Phi'_i}$ connected to the current space by nonzero Hamiltonian matrix elements $H_{ij} = \bra{\Phi'_i} \hat{H} \ket{\Phi_j}$ are added according to the importance measure
\be
   f(\ket{\Phi'_i}) = \max_j(|H_{ij} c_j|),
\ee
where $\{c_j\}$ are determinant coefficients. The variational stage is summarized as follows \cite{holmes_heat-bath_2016}:
\begin{enumerate}
    \item{
        Diagonalize the Hamiltonian in the selected space. Let $\{c_j\}$ denote the determinant coefficients of the lowest eigenvector.
    }

    \item{
        Generate all exterior determinants $\{\ket{\Phi'_i}\}$ that satisfy $|H_{ij} c_j| > \epsilon_1$ for at least one determinant $\ket{\Phi_j}$ in the current space. Add these determinants to the selected space.
    }

    \item{
        Repeat steps 1 and 2 until the number of new determinants generated is $<1\%$ of the number of determinants in the selected space.
    }
\end{enumerate}

The generation of $\{\ket{\Phi'_i}\}$ in step 2 is achieved via \emph{deterministic heat-bath sampling}. Starting from a reference determinant $\ket{\Phi_j}$, only determinants $\ket{\Phi'_i}$ connected to $\ket{\Phi_j}$ through matrix elements $\bra{\Phi'_i} \hat{H} \ket{\Phi_j}$ that satisfy $|\bra{\Phi'_i} \hat{H} \ket{\Phi_j}| > \epsilon$ are sampled. The threshold $\epsilon$ here is chosen to be $\epsilon = \epsilon_1 / |c_j|$. Duplicates are removed as the sampling algorithm is applied to each determinant in the current space. 

Once the variational wave function $\ket{\Psi_0} = \sum_i c_i \ket{\Phi_i}$ is obtained, the partitioning of the Hamiltonian used in the perturbative stage is defined as
\begin{gather}
    \hat{H} = \hat{H}_0 + \hat{V}, \nonumber \\
    \hat{H}_0 = \sum_{ij} H_{ij} \ket{\Phi_i} \bra{\Phi_j}, \quad \hat{H}_0 \ket{\Psi_0} = E_0 \ket{\Psi_0}.
\end{gather}
The second-order correction to the energy is thus
\begin{align}
    \Delta E_2 &= \sum_a \frac{\left( \sum_i H_{ai} c_i \right)^2}{E_0 - E_a}, \label{pt2_correction}
\end{align}
where the sum is over new determinants $\{\ket{\Phi'_a}\}$ that give the matrix elements $H_{ai} = \bra{\Phi'_a} \hat{H} \ket{\Phi_i}$ and $E_a = H_{aa}$. Since \eqref{pt2_correction} is expensive to evaluate, vHCI + PT2 approximates $\Delta E_2$ using only terms in the sum that satisfy $|H_{ai} c_i| > \epsilon_2$, \textit{i.e.}
\begin{align}
    \Delta E_2 \approx \sum_a \frac{\left( \sum_{i, |H_{ai} c_i| > \epsilon_2} H_{ai} c_i \right)^2}{E_0 - E_a}.
\end{align}
Similar to vHCI, the set $\{\ket{\Phi'_a}\}$ is generated by applying the heat-bath sampling algorithm to each determinant in $\ket{\Psi_0}$ using a threshold of $\epsilon = \epsilon_2/|c_i|$.

\subsection{Coupled-cluster theory}
CC theory \cite{shavitt_many-body_2009} assumes an ansatz for the $N$-particle ground state of the form
\be
    \ket{\Psi_\text{CC}} = e^{\hat{T}} \ket{\Psi_0}, \label{cc_ansatz}
\ee
where $\ket{\Psi_0}$ is a reference Slater determinant (often taken to be the Hartree-Fock solution). The \emph{cluster operator} $\hat{T}$ is given by
\begin{align}
    \hat{T} &= \hat{T}_1 + \hat{T}_2 + \cdots + \hat{T}_N, \\
    \hat{T}_1 &= \sum_{ia} t_i^a \hat{a}_a^\dagger \hat{a}_i, \\
    \hat{T}_2 &= \sum_{\substack{i<j\\a<b}} t_{ij}^{ab} \hat{a}_a^\dagger \hat{a}_b^\dagger  \hat{a}_i \hat{a}_j, \\
    &\qquad \vdots \nonumber
\end{align}
where $\hat{T}_i$ are cluster operators associated with $i$-particle excitations and $t_i^a, t_{ij}^{ab}, \dots$ are \emph{cluster amplitudes} to be determined. These amplitudes together with the CC energy can be computed through a projection of the Schr\"odinger equation onto $\ket{\Psi_0}$ and its excitations:
\begin{align}
    \bra{\Psi_0} e^{-\hat{T}} \left( \hat{H} - E\right) e^{\hat{T}} \ket{\Psi_0} &= 0, \\
    \bra{\Psi_0}\hat{a}_i^\dagger \hat{a}_a e^{-\hat{T}} \left( \hat{H} - E\right) e^{\hat{T}} \ket{\Psi_0} &= 0, \\
    \bra{\Psi_0}\hat{a}_j^\dagger \hat{a}_i^\dagger \hat{a}_b \hat{a}_a e^{-\hat{T}} \left( \hat{H} - E\right) e^{\hat{T}} \ket{\Psi_0} &= 0, \\
    &\ \ \vdots \nonumber
\end{align}
The resulting set of coupled polynomial equations (\textit{i.e.} the \emph{coupled-cluster equations}) can be solved (often iteratively) for the energy and amplitudes.

Note that $\ket{\Psi_\text{CC}}$ with $\hat{T}$ untruncated is simply a non-linear reparametrization of \eqref{ci_wfn}, \textit{i.e.} it is equivalent to exact diagonalization. The power of the CC approach, however, lies in the truncated form of the ansatz. The exponential form allows higher order excitations of $\ket{\Psi_0}$ to be included in the ansatz despite truncating $\hat{T}$. One of the most widely used variants is the truncation $\hat{T} = \hat{T}_1 + \hat{T}_2$, known as CCSD. The accuracy of CCSD can be na\"ively improved by including $\hat{T}_3$, but this approach is often too costly. A more feasible way to improve CCSD solutions is to include contributions from $\hat{T}_3$ perturbatively, leading to the CCSD(T) method which is often called the ``gold standard" of quantum chemistry.

\section{Computational details} \label{app:comp_details}
\subsection{Generalized Hartree-Fock}
The spinor structure of generalized Hartree-Fock (GHF) orbitals breaks $\hat{S}^2$ and $\hat{S}_z$ symmetry, hence $\ket{\Psi_\text{GHF}}$ is not constrained to be an eigenstate of $\hat{S}^2$ and $\hat{S}_z$. Sacrificing symmetries of the exact solution provides the additional flexibility necessary to obtain the true variational minimum, which can be advantageous in studying magnetically frustrated systems that require a non-spin-collinear description \cite{goings_stability_2015}. As previous work \cite{morales-duran_magnetism_2022} on TMD heterobilayers has suggested GWC ground states with non-collinear spins at some fractional filling factors, GHF is an appropriate mean-field method for our problem. 

We performed all GHF calculations using a development version of \texttt{Q-Chem} \cite{epifanovsky_software_2021}. Solutions were optimized through a hybrid scheme of first using the Direct Inversion in the Iterative Subspace (DIIS) method \cite{pulay_convergence_1980,pulay_improved_1982} for $\mathcal{O}(1)$ iterations before switching to the Geometric Direct Minimization (GDM) algorithm \cite{van_voorhis_geometric_2002}. This approach combines DIIS's ability to efficiently converge to the \emph{global minimum} during the early iterations together with GDM's ability to traverse challenging energy landscape features and robustly arrive at a \emph{local minimum}. We executed supercell calculations composed of $\sqrt{N_\text{cell}} \times \sqrt{N_\text{cell}}$ unit cells at the $\Gamma$ point of the mBZ, which is analogous to performing unit cell calculations at $N_k = N_\text{cell}$ $\vec{k}$-points adequately sampled from the mBZ. Periodic boundary conditions are only imposed across the supercell, which allow us to naturally obtain solutions where discrete translational symmetry on the moiré superlattice is broken. To break $C_3$ symmetry, however, requires explicit encoding of the broken symmetry in the initial guess. 

For our survey of GHF solutions across filling factors in Fig. \ref{fig:charge_gap} in the main text and Figs. \labelcref{fig:charge_density,fig:fft_peaks,fig:raw_fft_peaks}, we started with a ``noisy" spin-unpolarized initial guess constructed from continuum model orbitals, \textit{i.e.} eigenstates of \eqref{moire_h} in the main text (henceforth referred to as the ``continuum model guess"). By ``noisy", we mean that matrix elements of the spin $\uparrow \downarrow$ and $\downarrow \uparrow$ blocks of the density matrix $\mathbf{P}$ are randomly sampled from the standard Gaussian distribution $\mathcal{N}(0, 1)$, normalized by $10^{-3} \times \max |\mathbf{P}|$. At $\nu = 1/2$, we additionally considered a spin-polarized stripe guess constructed from orthogonal Bloch orbitals of the form
\be 
    \psi_{\vec{k}}(\vec{r}) \propto \sum_{i=1}^{N} e^{i \vec{k} \cdot \vec{R}_i} \phi_i(\vec{r}),
\ee 
where $N$ is the particle number, $\phi_i(\vec{r})$ are Gaussians centered at stripe lattice sites $\vec{R}_i$, and $\vec{k}$ are crystal momenta that:
\begin{enumerate}[label=(\roman*)]
    \item{
        are restricted to the first Brillouin zone defined by the reciprocal space
        of the stripe lattice;
    }

    \item{
        satisfy the periodic boundary conditions imposed on the supercell.
    }
\end{enumerate}

To evaluate the basis set convergence of solutions, we performed a series of calculations with increasing basis set size $N_\text{basis}$ starting from a small $N_\text{basis}^\text{min}$. The converged solution at each $N_\text{basis}$ was input as the initial guess for the next calculation using a larger $N_\text{basis}$, where the $N_\text{basis}^\text{min}$ solutions were first ensured to be stable.

\subsection{Stability analysis}
The self-consistent procedure used in GHF only guarantees solutions to be stationary points. To determine whether they are local minima or saddle points, we inspect the lowest eigenvalues of the electronic Hessian following the approach of \cite{sharada_wavefunction_2015}. Diagonalization is executed via a finite-difference implementation of the Davidson method, where for a vector $\vec{v}$ in the subspace, the Hessian-vector product is computed as
\be
\mathbf{H} \vec{v} \approx \left[\frac{\nabla E(\mathbf{C}_{+}) - \nabla E(\mathbf{C}_{-})}{2\xi} \right]^\mathsf{T}.
\ee
$\nabla E$ is the analytic energy gradient, while $\mathbf{C}_{\pm}$ are the GHF solution coefficient matrices perturbed in opposite directions determined by the finite step size $\xi$. If the solution is deemed unstable, we can displace the solution along the lowest eigenvector and use this as an initial guess in a new GHF calculation.

\subsection{Spin-collinearity test}
We evaluated the spin collinearity of our GHF solutions following the approach of \cite{small_simple_2015}. A wave function $\ket{\Psi}$ is considered to be collinear if it is an eigenstate of an axial spin operator $\hat{S}_c = \vec{c} \cdot \hat{\vec{S}}$ associated with some Cartesian coordinate, where $\vec{c} \in \mathbb{R}^3$ is a unit vector and $\hat{\vec{S}} = (\hat{S}_x, \hat{S}_y, \hat{S}_z)$ is the usual many-body spin vector operator. In other words, $\ket{\Psi}$ is collinear if it satisfies the relation 
\begin{gather}
    \hat{S}_c \ket{\Psi} = s_c \ket{\Psi}, \\
    s_c \in \{-N/2, -N/2 + 1, \dots, N/2-1, N/2\}. \nonumber
\end{gather}
The collinearity test for a state consists of computing the quantities $\mu$ and $\varepsilon_0$, defined as
\begin{align}
    \mu &= \langle (\hat{S}_c)^2 \rangle - \langle \hat{S}_c \rangle^2, \\
    \varepsilon_0 &= \norm{\langle \hat{\vec{S}} \rangle} = \norm{\left(\langle \hat{S}_x \rangle, \langle \hat{S}_y \rangle, \langle \hat{S}_z \rangle \right)},
\end{align}
where $\norm{\cdot}$ denotes the Euclidean norm. The conditions for collinearity are thus:
\begin{enumerate}[label=(\roman*)]
    \item{$\mu = 0$.}
    \item{$\varepsilon_0 \in \{-N/2, -N/2 + 1, \dots, N/2-1, N/2\}.$}
\end{enumerate}
Condition (i) is obvious since it can easily be proved that $\mu = 0$ if and only if $\ket{\Psi}$ is an eigenstate of $\hat{S}_c$. Condition (ii) is less straightforward and we refer the reader to \cite{small_simple_2015} for its justification. In practice, we deem $\mu \lesssim 10^{-4}$ as describing a collinear state.

\subsection{Correlated methods}
We performed correlated calculations using only spin-collinear variants of HCI, CCSD, and CCSD (T). CCSD and CCSD(T) calculations were carried out using \texttt{PySCF} \cite{sun_libcint_2015,sun_pyscf_2018,sun_recent_2020}, whereas HCI calculations with $\sim \mathcal{O}(10^{6})$ determinants and parameters $\epsilon_1, \epsilon_2 \approx 10^{-7}$ were executed using a locally modified version of \texttt{Dice} \cite{holmes_heat-bath_2016,sharma_semistochastic_2017,smith_cheap_2017}. 

Recall that our HF calculations utilized a complex plane wave basis $\left\{ e^{i\vec{G}_p \cdot \vec{r}} \right\}$ which yields two-electron integral tensors $\braket{pq}{rs}$ possessing a fourfold symmetry, \textit{i.e.}
\begin{align}
    \braket{pq}{rs} &= \braket{rs}{pq} = \braket{sr}{qp} = \braket{qp}{sr}.
\end{align}
Quantum chemistry codes including \texttt{PySCF}, however, often assume the basis sets to be real and hence $\braket{pq}{rs}$ to be eightfold symmetric, \textit{i.e.}
\begin{align}
    \braket{pq}{rs} &= \braket{rq}{ps} = \braket{ps}{rq} = \braket{rs}{pq} \nonumber \\
    = \braket{qp}{sr} &= \braket{sp}{qr} = \braket{qr}{sp} = \braket{sr}{qp}.
\end{align}
Thus, to interface with such codes, we applied a unitary transformation to rotate the plane wave basis into a real-valued basis composed of sines and cosines (up to a normalization factor):
\begin{align}
    \begin{bmatrix}
        e^{i\vec{G}_p \cdot \vec{r}} \\ e^{-i\vec{G}_p \cdot \vec{r}}
    \end{bmatrix}
    &\xlongrightarrow{\quad \mathbf{U} \quad}
    \sqrt{2}
    \begin{bmatrix}
        \cos{(\vec{G}_p \cdot \vec{r})} \\ \sin{(\vec{G}_p \cdot \vec{r})}
    \end{bmatrix}, \\[1em]
    \mathbf{U} &= 
    \frac{1}{\sqrt{2}}
    \begin{bmatrix}
    1 & -i \\ 1 & i 
    \end{bmatrix}.
\end{align}
The resulting two-electron integral tensor has eightfold symmetry recovered.

\onecolumngrid
\section{FFT of the charge density} \label{app:fft_raw}
\begin{figure*}[!b]
    \centering
    \includegraphics[width=0.85\textwidth]{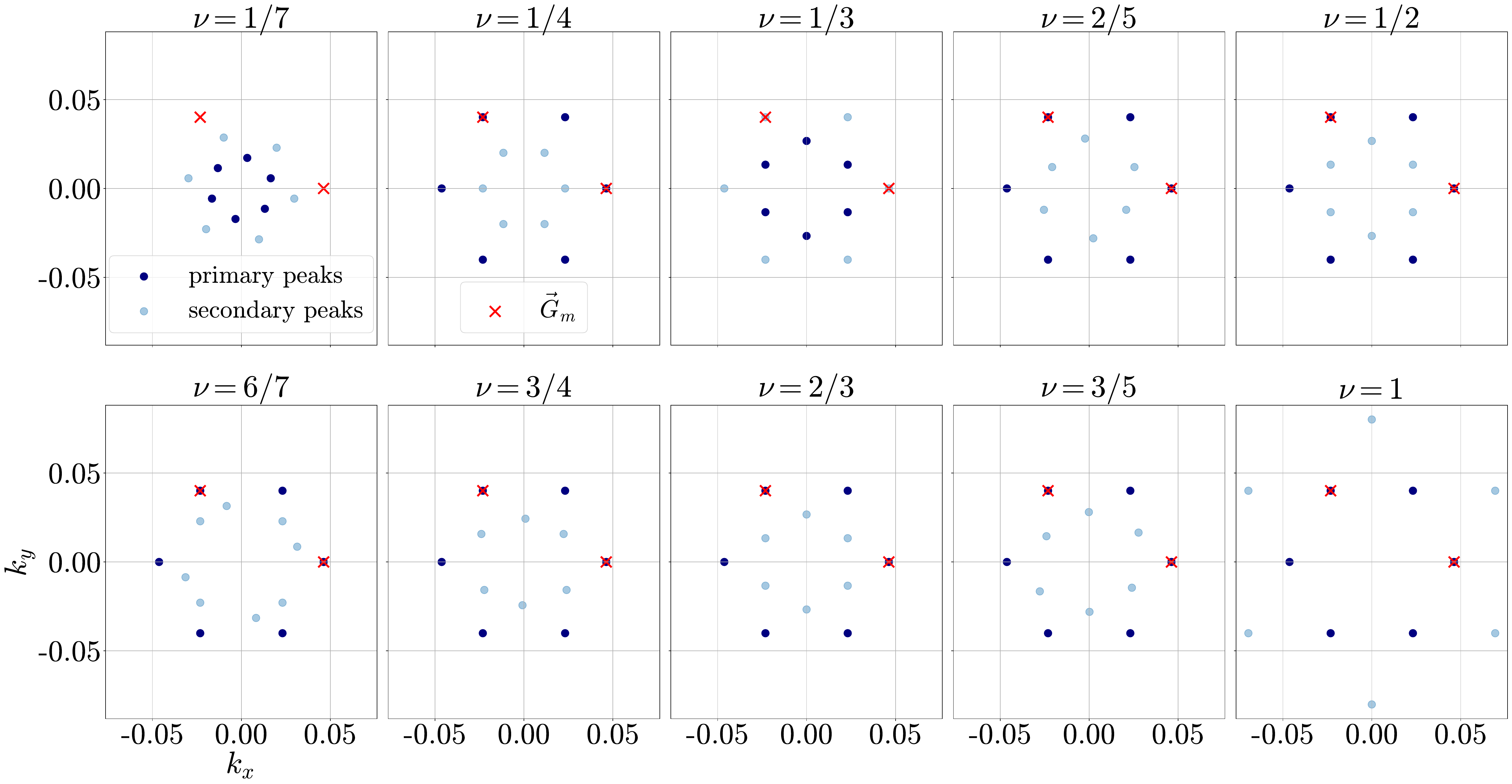}
    \caption{Peaks extracted from Fig. \ref{fig:raw_fft_peaks}. We use the term ``primary" to describe peaks with the largest magnitudes, while ``secondary" describes any feature of secondary importance. $\vec{G}_m$ denotes the primitive moiré reciprocal vector.}
    \label{fig:fft_peaks}
\end{figure*}

\begin{figure}[!tb]
    \centering
    \begin{subfigure}[b]{0.23\textwidth}
        \centering
        \includegraphics[width=\textwidth]{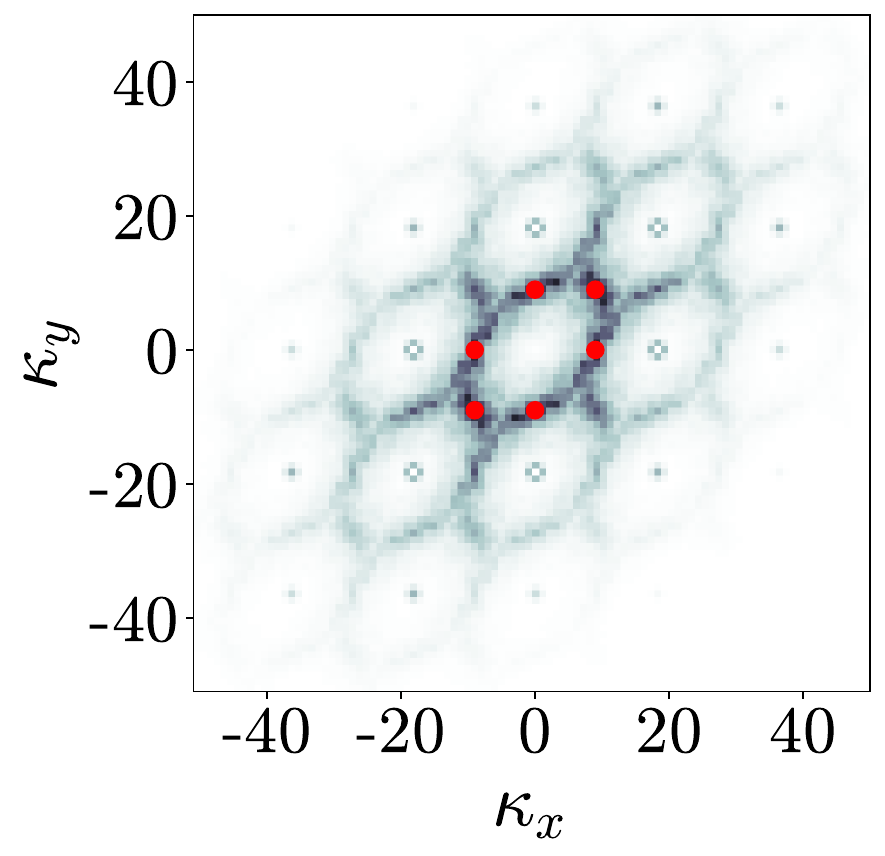}
        \caption{$\nu = 1/4$}
        \label{fig:extract_charge_gap_nu=14}    
    \end{subfigure}
    \begin{subfigure}[b]{0.24\textwidth}
        \centering
        \includegraphics[width=\textwidth]{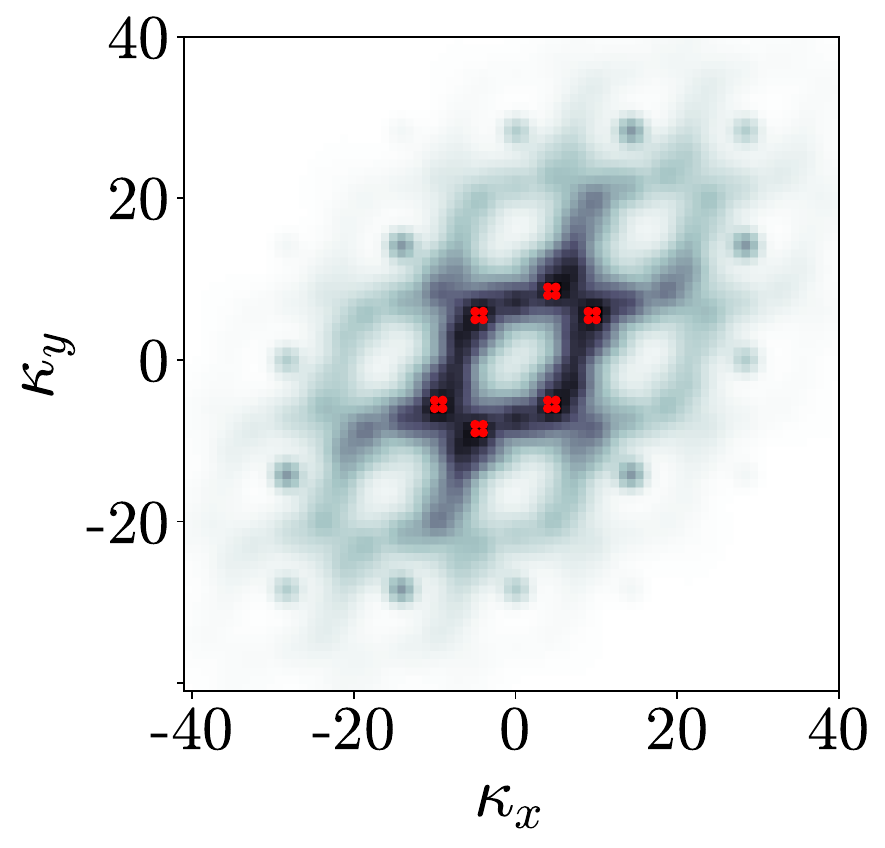}
        \caption{$\nu = 3/4$}
        \label{fig:extract_charge_gap_nu=34}    
    \end{subfigure}
    \caption{Illustration of our FFT secondary peak extraction by: (i) removing the primary peaks at the moiré reciprocal lattice sites; (ii) applying a Gaussian blur filter to smooth out $|\rho(\vec{k})|$ near the secondary peaks; and (iii) extracting the locations where $|\rho(\vec{k})|$ is above some threshold value directly or via $k$-means clustering.}
    \label{fig:extract_charge_gap}
\end{figure}

Prior to computing the FFT of the charge densities $\rho(\vec{r})$ in Fig. \ref{fig:charge_density}, we subtracted the average density $\langle \rho \rangle$ to remove the $\vec{k} = 0$ Fourier component. To yield better insight into the periodic structure of the charge densities in Fig. \ref{fig:charge_density}, we extracted the locations where the FFT signal was maximum (\textit{i.e.} primary) and where the signal was sub-maximum (\textit{i.e.} secondary). The results are shown in Fig. \ref{fig:fft_peaks}. At $\nu = 1/7, \ 1/3, \ 2/3, \ 6/7,$ and $1$, the FFT exhibits clean peaks that can be easily extracted. At $\nu = 1/4, \ 2/5, \ 1/2, \ 3/5,$ and $3/4$, however, in addition to the clear peaks located at the moiré reciprocal lattice sites, the FFT displays secondary disperse features in reciprocal space associated with charge delocalization in real space. We estimated the locations of possible secondary peaks suppressed by the disperse features but nonetheless can still be gauged by eye. An example is shown for $\nu = 1/4$ in Fig. \ref{fig:extract_charge_gap_nu=14}, where we first removed the primary peaks at the moiré reciprocal lattice sites before applying a Gaussian blur filter to smooth out $|\rho(\vec{k})|$ near the secondary peaks. The locations (marked by the red dots) where $|\rho(\vec{k})|$ is above some threshold value are then extracted. In cases where there are clusters of similar $|\rho(\vec{k})|$ above the threshold value, as for $\nu = 3/4$ in Fig. \ref{fig:extract_charge_gap_nu=34}, we identified the cluster center by $k$-means clustering.

\section{Finite-size effects} \label{app:finite-size errors}
\begin{figure*}[!ht]
    \centering
    \includegraphics[width=0.68\textwidth]{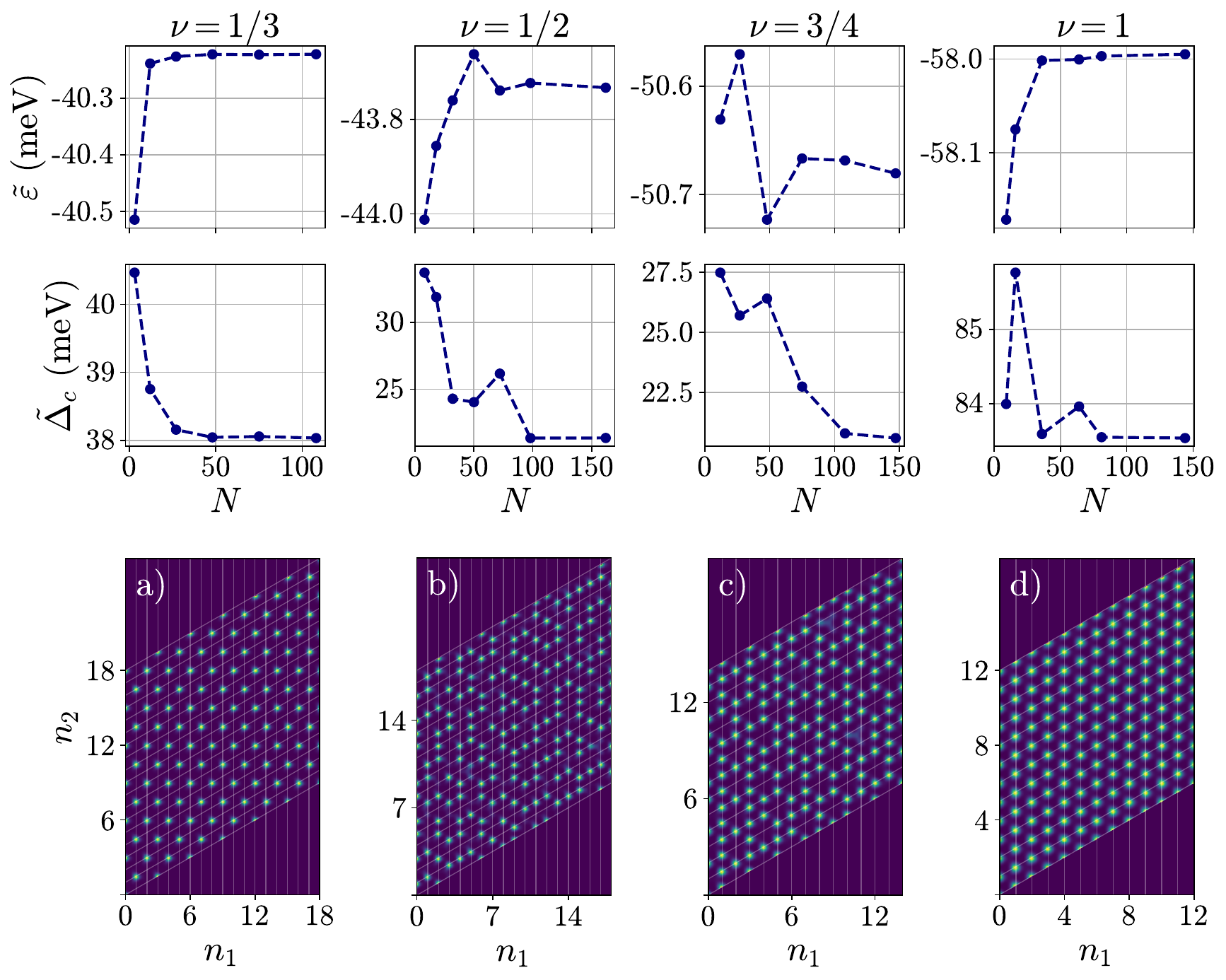}
    \caption{Convergence of the Madelung-corrected total energy per particle $\tilde\varepsilon$ and charge gap $\tilde\Delta_c$ with system size $N$. The values at $\nu = 1/3$ and 1 were derived from ordered phases, while those at $\nu = 1/2$ and 3/4 were derived from disordered phases, as depicted in panels (a)-(d).}
    \label{fig:fs_extrap}
\end{figure*}

For a $N$-particle system, the Madelung-corrected total energy per particle $\varepsilon(N) = E(N)/N$ and charge gap $\Delta_c(N) = E(N+1) + E(N-1) - 2E(N)$ are computed as \cite{yang_electronic_2020,xing_unified_2021}
\begin{align}
    \tilde{\varepsilon}(N) &= \varepsilon(N) + \varepsilon_M \label{app:fs_corrected_energy}, \\
    \tilde{\Delta}_c(N) &= \Delta_c(N) + 2|\varepsilon_M| \label{app:fs_corrected_charge_gap},
\end{align}
which improves the scaling in the finite-size errors from $\mathcal{O}(N^{-1/2})$ to $\mathcal{O}(N^{-1})$ for $\varepsilon$ \cite{xing_unified_2021} and to $\mathcal{O}(N^{-3/2})$ for $\Delta_c$ \cite{hunt_quantum_2018}. Our extrapolations of $\tilde{\varepsilon}$ and $\tilde{\Delta}_c$ to $N \to \infty$ thus assume $\mathcal{O}(N^{-1})$ and $\mathcal{O}(N^{-3/2})$ scalings, respectively. As illustrative examples, Fig. \ref{fig:fs_extrap} shows the convergence of $\tilde{\varepsilon}$ and $\tilde{\Delta}_c$ with $N$ at $\nu = 1/3, 1/2, 3/4,$ and $1$. It should be noted that the values at $\nu = 1/3$ and 1 were derived from ordered phases, while those at $\nu = 1/2$ and 3/4 were derived from disordered phases, as depicted in Figs. \ref{fig:fs_extrap}a-d. This likely contributes to the more monotonic convergence in $\tilde{\varepsilon}$ and $\tilde{\Delta}_c$ at $\nu = 1/3$ and 1 compared to $\nu = 1/2$ and 3/4. For the latter case, the greater variation may be associated with disordered configurations that differ slightly from one another. The variation in $\tilde{\varepsilon}$ as a function of $N$ is also generally smaller than in $\tilde{\Delta}_c$. 

\begin{figure*}[!h]
    \centering
    \includegraphics[width=0.48\textwidth]{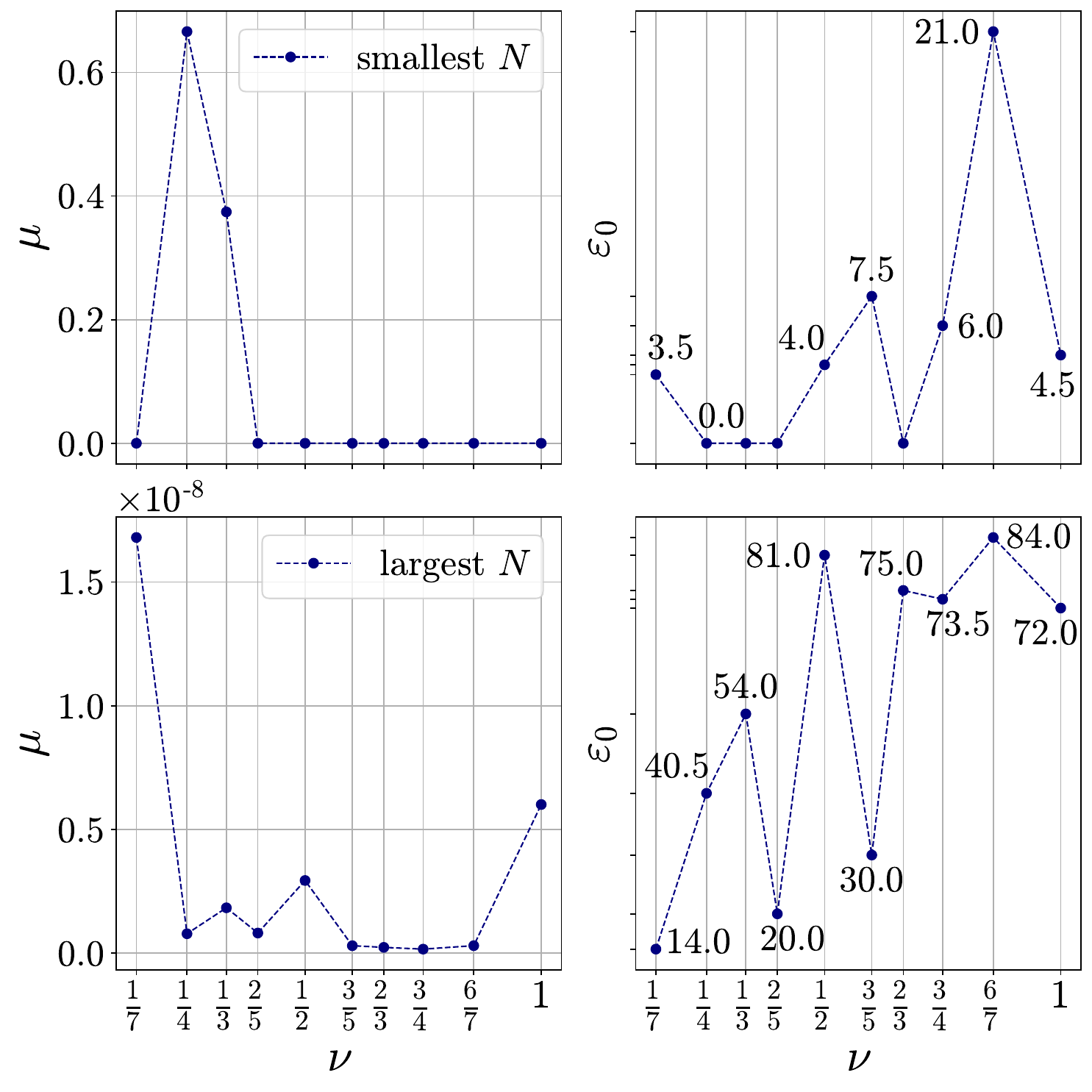}
    \caption{The quantities $\mu$ and $\varepsilon_0$ of GHF solutions initialized from the unpolarized continuum model guess. While the smallest $N$-particle system we studied at each $\nu$ may not be collinear, the largest $N$-particle system is always collinear. This suggests that the emergence of non-collinear solutions could be an artifact of finite-size effects. In the order of increasing $\nu$, the smallest (largest) $N$ we considered were 7 (28), 4 (81), 3 (108), 10 (40), 8 (162), 15 (60), 6 (150), 12 (147), 42 (168), and 9 (144).}
    \label{fig:spin_collinearity}
\end{figure*}

Furthermore, a manifestation of finite-size effects was observed in the spin-collinearity of solutions initialized from the continuum model guess, as shown in Fig. \ref{fig:spin_collinearity}. While only collinear polarized solutions were realized for the largest system size studied at each $\nu$, the smallest system may be non-collinear. This suggests that--at least at the Hartree-Fock level--the spin properties of low-lying states may be sensitive to finite-size effects. However, we note that our initial guess was occasionally found to be insufficient in biasing solutions away from ferromagnetic behavior; lower energy solutions may indeed be non-collinear. As the primary goal of this paper is to examine charge orders, we do not believe any of our main conclusions are affected.


\section{Screened Coulomb potential} \label{app:screened_coulomb}
\begin{figure*}[!h]
    \centering
    \includegraphics[width=0.7\textwidth]{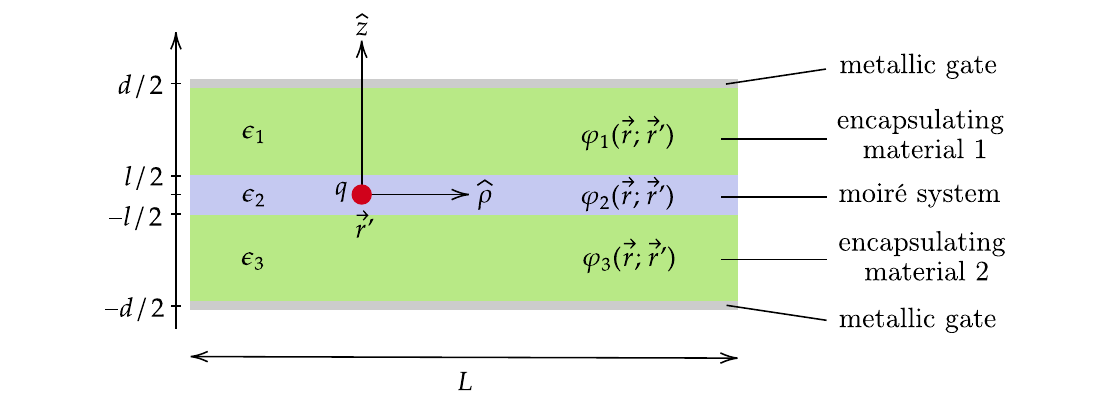}
    \caption{The setup for deriving the screened Coulomb potential.}
    \label{fig:screened_coulomb_setup}
\end{figure*}

In experiments on TMD moiré systems that we consider, the setup often involves encapsulating the moiré system between dielectric materials and metallic gates (Fig. \ref{fig:screened_coulomb_setup}). The encapsulating material and gates alter the electrostatic environment for a charge in the TMD layer, resulting in a screened Coulomb interaction between in-plane charges. Here, we derive a screened Coulomb potential that is a generalization of the double gate-screened \cite{morales-duran_magnetism_2022,faulstich_interacting_2022,bernevig_twisted_2021} and Rytova-Keldysh \cite{rytova_screened_1967,keldysh_coulomb_1979} potentials considered in moiré and TMD materials, respectively. 

Regarding the setup in Fig. \ref{fig:screened_coulomb_setup}, our task is to find the electric potential $\varphi_2(\vec{r}; \pvec{r}')$ at a point $\vec{r} = (x, y, z)$ due to a charge $q$ located at $\pvec{r}' = (0, 0, z')$, where $z, z' \in [-l/2, l/2]$. Let $\varphi_i$ and $\epsilon_i$ denote the potential and dielectric constant in layer $i$, respectively. Specializing to the case where $\epsilon_1 = \epsilon_3$, the electrostatic equations to solve are
\begin{align}
    \nabla_{\Vec{r}}^2 \ \varphi_1(\Vec{r}; \pvec{r}') &= 0, \label{em1} \\ 
    \nabla_{\Vec{r}}^2 \ \varphi_2(\Vec{r}; \pvec{r}') &= -\frac{4\pi q}{\epsilon_2} \delta(\Vec{r} - \pvec{r}'), \label{em2} \\ 
    \nabla_{\Vec{r}}^2 \ \varphi_3(\Vec{r}; \pvec{r}') &= 0, \label{em3}
\end{align}
with BCs (dropping the $\pvec{r}'$ parameter for notational clarity)
\begin{align}
    \varphi_1 \left(\Vec{\rho}, \frac{d}{2}\right) &= 0, \label{bc1} \\
    \varphi_1 \left(\Vec{\rho}, \frac{l}{2}\right) &= \varphi_2 \left(\Vec{\rho}, \frac{l}{2}\right) \label{bc2}, \\
    \varphi_2 \left(\Vec{\rho}, -\frac{l}{2}\right) &= \varphi_3 \left(\Vec{\rho}, -\frac{l}{2}\right) \label{bc3}, \\
    \varphi_3 \left(\Vec{\rho}, -\frac{d}{2}\right) &= 0, \label{bc4} 
\end{align}
\begin{align}
    \epsilon_1 \frac{\partial \varphi_1}{\partial z} \left(\Vec{\rho}, \frac{l}{2}\right) &= \epsilon_2 \frac{\partial \varphi_2}{\partial z} \left(\Vec{\rho}, \frac{l}{2}\right) \label{bc5}, \\
    \epsilon_2 \frac{\partial \varphi_2}{\partial z} \left(\Vec{\rho}, -\frac{l}{2}\right) &= \epsilon_1 \frac{\partial \varphi_3}{\partial z} \left(\Vec{\rho}, -\frac{l}{2}\right). \label{bc6}
\end{align}
$\vec{\rho} \in \mathbb{R}^2$ is a vector in the $xy$-plane. Furthermore, we assume $d \ll l$ so we can treat the layers as possessing infinite extent in the $xy$-plane. Fourier transforming \eqref{em1}-\eqref{em3} in $\vec{\rho}$, we obtain a system of second-order partial differential equations (PDEs)
\begin{align}
    \frac{\partial^2}{\partial z^2} \varphi_1(\vec{k}, z) - k^2 \varphi_1(\vec{k}, z) &= 0, \\
    \frac{\partial^2}{\partial z^2} \varphi_2(\vec{k}, z) - k^2 \varphi_2(\vec{k}, z) &= -\frac{4\pi q}{\epsilon_2} \delta(z - z'), \\
    \frac{\partial^2}{\partial z^2} \varphi_3(\vec{k}, z) - k^2 \varphi_3(\vec{k}, z) &= 0.
\end{align}

Let us define the operator $\mathcal{L} = \frac{\partial^2}{\partial z^2} - k^2$. Since $\varphi_2(\vec{k}, z)$ is the solution to an \emph{inhomogeneous} PDE with \emph{inhomogeneous} BCs, we can write it as
\begin{align}
    \varphi_2(\vec{k}, z) &= \tilde{\varphi}_2(\vec{k}, z) + G(\vec{k}, z),
\end{align}
where $\tilde{\varphi}_2(\vec{k}, z)$ is the solution to the \emph{homogeneous} PDE $\mathcal{L} \varphi(\vec{k}, z) = 0$ with \emph{inhomogeneous} BCs, while $G(\vec{k}, z)$ is the solution to the \emph{inhomogeneous} PDE $\mathcal{L} \varphi(\vec{k}, z) = -(4\pi q/\epsilon_2) \delta(z - z')$ with \emph{homogeneous} BCs. The screened potential we obtain from solving the system of PDEs is given by

\begin{align}
        &\tilde{\varphi}_2(\vec{k}, z; \pvec{r}') \nonumber \\
        &= \Xi \Bigg\{ \left[ e^{k(z+l/2)} \sinh{\left[ k \left( \frac{l}{2} + z' \right) \right]} + e^{-k(z-l/2)} \sinh{\left[ k \left( \frac{l}{2} - z' \right) \right]} \right] \left[ \frac{\epsilon_2}{\epsilon_1} \left( 1 - \cosh{[k(d-l)]} \right) - \sinh{[k(d-l)]} \right]  \nonumber \\
        &\qquad + \left[ e^{k(z-l/2)} \sinh{\left[ k \left( \frac{l}{2} - z' \right) \right]} + e^{-k(z+l/2)} \sinh{\left[ k \left(\frac{l}{2} + z' \right) \right]} \right] \left[ \frac{\epsilon_2}{\epsilon_1} \left( 1 - \cosh{[k(d-l)]} \right) + \sinh{[k(d-l)]} \right] \Bigg\}, \label{tilde_varphi2}
\end{align}
\begin{align}
    \Xi &= -\frac{4 \pi q}{\epsilon_1 k \sinh{(kl)}} \left\{ \left(\frac{\epsilon_2}{\epsilon_1} + 1 \right)^2 \sinh{(kd)} - \left(\frac{\epsilon_2}{\epsilon_1} - 1 \right)^2 \sinh{[k(d-2l)]} - 2 \left[ \left(\frac{\epsilon_2}{\epsilon_1} \right)^2 - 1 \right] \sinh{(kl)} \right\}^{-1}, \label{xi_varphi2}
\end{align}
\begin{align}
    G(\vec{k}, z; \pvec{r}') &= -\frac{2 \pi q}{\epsilon_2 k \sinh{(kl)}} \left\{ \cosh{[k(z+z')]} - \cosh{[k\left(|z-z'|-l\right)]} \right\}. \label{g_varphi2}
\end{align}

\subsection{Double gate-screened potential}
In the limit $\epsilon_1 = \epsilon_2$, the BCs match those of the double gate-screened potential and so we recover this solution. Setting $\epsilon_1 = \epsilon_2$ in \eqref{tilde_varphi2}-\eqref{g_varphi2},

\begin{align}
        &\tilde{\varphi}_2(\vec{k}, z; \pvec{r}') \nonumber \\
        &= \Xi \Bigg\{ \left[ e^{k(z+l/2)} \sinh{\left[ k \left( \frac{l}{2} + z' \right) \right]} + e^{-k(z-l/2)} \sinh{\left[ k \left( \frac{l}{2} - z' \right) \right]} \right] \left[ 1 - \cosh{[k(d-l)]} - \sinh{[k(d-l)]} \right]  \nonumber \\
        &\qquad + \left[ e^{k(z-l/2)} \sinh{\left[ k \left( \frac{l}{2} - z' \right) \right]} + e^{-k(z+l/2)} \sinh{\left[ k \left(\frac{l}{2} + z' \right) \right]} \right] \left[ 1 - \cosh{[k(d-l)]} + \sinh{[k(d-l)]} \right] \Bigg\},
\end{align}
\begin{align}
    \Xi &= -\frac{4 \pi q}{\epsilon k \sinh{(kl)}} \left\{ 4 \sinh{(kd)} \right\}^{-1},
\end{align}
\begin{align}
    G(\vec{k}, z; \pvec{r}') &= -\frac{2 \pi q}{\epsilon k \sinh{(kl)}} \left\{ \cosh{[k(z+z')]} - \cosh{[k\left(|z-z'|-l\right)]} \right\}.
\end{align}

Since Coulomb interactions in the GWC states occur over the length scale of the moiré lattice constant $L_m \gg l$, we are interested only in the asymptotic behaviour of $\varphi_2(\vec{r}; \pvec{r}')$ for distances $\rho \gg l$. Using the fact that $k \propto 1/\rho \ll 1/l$ in this case, it is sufficient for us to know the Fourier component $\varphi_2(\vec{k}, z; \pvec{r}')$ for $kl \ll 1$. It then follows that $k|z|, k|z'| \ll 1$ and $k|z-z'| \approx 0$ since $-l/2 \leq z, z' \leq l/2$. Taylor expanding $\tilde{\varphi}_2$ and $G$ to first order in $kl$ and making the approximation $z \approx z'$ (note that we still have $k|z| \sim \mathcal{O}(kl)$), we obtain

\begin{align}
    \tilde{\varphi}_2(\Vec{k}, z)
    &\approx \Xi \left\{ 2kl (1 - \cosh{kd}) + (kl)^2 \sinh{kd} - 4(kz)^2 \sinh{kd} \right\} \nonumber \\
    &= -\frac{\pi q}{\epsilon k \sinh{kd}} \left\{ 2(1 - \cosh{kd}) + kl \sinh{kd} - 4\frac{(kz)^2}{kl} \sinh{kd} \right\}, \\
    G(\vec{k}, z) &\approx -\frac{2 \pi q}{\epsilon k} \left[ 2\frac{(kz)^2}{kl} - \frac{1}{2} kl \right],
\end{align}
\begin{align}
    \implies \varphi_2(\vec{k}, z) &\approx -\frac{\pi q}{\epsilon k \sinh{kd}} \left\{ 2(1 - \cosh{kd}) + \cancel{kl \sinh{kd}} - \cancel{4\frac{(kz)^2}{kl} \sinh{kd}} + \cancel{4\frac{(kz)^2}{kl} \sinh{kd}} - \cancel{kl \sinh{kd}} \right\} \nonumber \\
    &= \frac{2 \pi q}{\epsilon k} \cdot \tanh{\left(\frac{kd}{2}\right)}.
\end{align}

\subsection{Rytova-Keldysh potential}
In the limit $d \to \infty$, the BCs are similar to the setup in \cite{rytova_screened_1967} and we recover Eq. (5) therein. We first shift our coordinates $z \to z - l/2$ and $z' \to z' - l/2$ in \eqref{tilde_varphi2}-\eqref{g_varphi2} to match the coordinate system of \cite{rytova_screened_1967}:

\begin{align}
        \tilde{\varphi}_2(\vec{k}, z; \pvec{r}') &= \Xi \Bigg\{ \left[ e^{kz} \sinh{(kz')} - e^{-k(z-l)} \sinh{\left[ k(z' - l) \right]} \right] \left[ \frac{\epsilon_2}{\epsilon_1} \left( 1 - \cosh{[k(d-l)]} \right) - \sinh{[k(d-l)]} \right]  \nonumber \\
        &\qquad + \left[ e^{-kz} \sinh{(kz')} - e^{k(z-l)} \sinh{\left[ k(z' - l) \right]} \right] \left[ \frac{\epsilon_2}{\epsilon_1} \left( 1 - \cosh{[k(d-l)]} \right) + \sinh{[k(d-l)]} \right] \Bigg\},
\end{align}
\begin{align}
    \Xi &= -\frac{4 \pi q}{\epsilon_1 k \sinh{(kl)}} \left\{ \left(\frac{\epsilon_2}{\epsilon_1} + 1 \right)^2 \sinh{(kd)} - \left(\frac{\epsilon_2}{\epsilon_1} - 1 \right)^2 \sinh{[k(d-2l)]} - 2 \left[ \left(\frac{\epsilon_2}{\epsilon_1} \right)^2 - 1 \right] \sinh{(kl)} \right\}^{-1},
\end{align}
\begin{align}
    G(\vec{k}, z; \pvec{r}') &= -\frac{2 \pi q}{\epsilon_2 k \sinh{(kl)}} \left\{ \cosh{[k(z+z'-l)]} - \cosh{[k\left(|z-z'|-l\right)]} \right\}.
\end{align}

Applying the hyperbolic relations
\begin{align}
    \cosh{[k(d-l)]} &= \cosh{(kd)}\cosh{(kl)} - \sinh{(kd)}\sinh{(kl)}, \\
    \sinh{[k(d-l)]} &= \sinh{(kd)}\cosh{(kl)} - \cosh{(kd)}\sinh{(kl)}, \\
    \sinh{[k(d-2l)]} &= \sinh{(kd)}\cosh{(2kl)} - \cosh{(kd)}\sinh{(2kl)},
\end{align}
and rewriting in terms of
\begin{align}
    \delta &= \frac{\epsilon_2/\epsilon_1 - 1}{\epsilon_2/\epsilon_1 + 1},
\end{align}
 we have upon taking $d \to \infty$:
 
\begin{align}
    \tilde{\varphi}_2(\vec{k}, z; \pvec{r}') &\approx \frac{2\pi q}{\epsilon_2 k \sinh{(kl)}} \cdot \frac{1 + \delta}{e^{2kl} - \delta^2} \cdot e^{kl} \Big\{ e^{kl} \cosh{[k(z+z'-l)]} - \cosh{[k(z-z')]} \nonumber \\
    &\qquad \qquad \qquad \qquad \qquad \qquad \qquad + \delta [\cosh{k(z-z')}] - e^{-kl} \cosh{[k(z+z'-l)]} \Big\}, \\
    \implies \varphi_2(\vec{k}, z; \pvec{r}') &= \tilde{\varphi}_2(\vec{k}, z; \pvec{r}') + G(\vec{k}, z; \pvec{r}') \nonumber \\
    &\approx \frac{2\pi q}{\epsilon_2 k} \left\{ e^{-k|z-z'|} + \frac{2\delta}{e^{2kl} - \delta^2} \left[ \delta \cosh{[k(z-z')]} + e^{kl} \cosh[k(z+z'-l)] \right] \right\}.
\end{align}

\end{document}